\documentclass[12pt]{article}
\usepackage{amssymb,graphicx} 

\begin{document}

\newcommand{\sect}[1]{\setcounter{equation}{0}\section{#1}}
\renewcommand{\theequation}{\thesection.\arabic{equation}}
\newcommand{\be}{\begin{equation}}
\newcommand{\ee}{\end{equation}}
\newcommand{\bea}{\begin{eqnarray}}
\newcommand{\eea}{\end{eqnarray}}
\newcommand{\eps}{\epsilon}
\newcommand{\om}{\omega}
\newcommand{\vph}{\varphi}
\newcommand{\sig}{\sigma}
\newcommand{\CC}{\mbox{${\mathbb C}$}}
\newcommand{\RR}{\mbox{${\mathbb R}$}}
\newcommand{\QQ}{\mbox{${\mathbb Q}$}}
\newcommand{\ZZ}{\mbox{${\mathbb Z}$}}
\newcommand{\NN}{\mbox{${\mathbb N}$}}
\newcommand{\1}{\mbox{\hspace{.0em}1\hspace{-.24em}I}}
\newcommand{\II}{\mbox{${\mathbb I}$}}
\newcommand{\prt}{\partial}
\newcommand{\und}[1]{\underline{#1}}
\newcommand{\wh}[1]{\widehat{#1}}
\newcommand{\wt}[1]{\widetilde{#1}}
\newcommand{\mb}[1]{\ \mbox{\ #1\ }\ }
\newcommand{\half}{\frac{1}{2}}
\newcommand{\noin}{\not\!\in}
\newcommand{\rhotimes}{\mbox{\raisebox{-1.2ex}{$\stackrel{\displaystyle\otimes}
{\mbox{\scriptsize{$\rho$}}}$}}}
\newcommand{\bin}[2]{{\left( {#1 \atop #2} \right)}}
\newcommand{\ri}{{\rm i}}
\newcommand{\rd}{{\rm d}}
\newcommand{\A}{{\cal A}}
\newcommand{\B}{{\cal B}}
\newcommand{\C}{{\cal C}}
\newcommand{\F}{{\cal F}}
\newcommand{\E}{{\cal E}}
\newcommand{\cP}{{\cal P}}
\newcommand{\R}{{\cal R}}
\newcommand{\T}{{\cal T}}
\newcommand{\W}{{\cal W}}
\newcommand{\cS}{{\cal S}}
\newcommand{\bS}{{\bf S}}
\newcommand{\cL}{{\cal L}}
\newcommand{\hlp}{{\RR}_+}
\newcommand{\hlm}{{\RR}_-}
\newcommand{\Hil}{{\cal H}}
\newcommand{\D}{{\cal D}}
\newcommand{\G}{{\cal G}}
\newcommand{\alg}{\C} 
\newcommand{\rep}{\F(\A)}
\newcommand{\trep}{\G_\beta(\A)}
\newcommand{\form}{\langle \, \cdot \, , \, \cdot \, \rangle }
\newcommand{\e}{{\rm e}}
\newcommand{\by}{{\bf y}}
\newcommand{\bp}{{\bf p}}
\newcommand{\LL}{\mbox{${\mathbb L}$}}
\newcommand{\Rp}{{R^+_{\, \, \, \, }}}
\newcommand{\Rm}{{R^-_{\, \, \, \, }}}
\newcommand{\Rpm}{{R^\pm_{\, \, \, \, }}}
\newcommand{\Tp}{{T^+_{\, \, \, \, }}}
\newcommand{\Tm}{{T^-_{\, \, \, \, }}}
\newcommand{\Tpm}{{T^\pm_{\, \, \, \, }}}
\newcommand{\baral}{\bar{\alpha}}
\newcommand{\barbt}{\bar{\beta}}
\newcommand{\supp}{{\rm supp}\, }
\newcommand{\Pt}{\widetilde{P}}
\newcommand{\At}{\widetilde{A}}
\newcommand{\Bt}{\widetilde{B}}
\newcommand{\St}{\widetilde{S}}
\newcommand{\kt}{\widetilde{k}}
\newcommand{\EE}{\mbox{${\mathbb E}$}}
\newcommand{\JJ}{\mbox{${\mathbb J}$}}
\newcommand{\MM}{\mbox{${\mathbb M}$}}
\newcommand{\ct}{{\cal T}}
\newcommand{\ph}{\varphi}
\newcommand{\phd}{\widetilde{\varphi}}
\newcommand{\phl}{\varphi_{{}_L}}
\newcommand{\phr}{\varphi_{{}_R}}
\newcommand{\phpl}{\varphi_{{}_{+L}}}
\newcommand{\phpr}{\varphi_{{}_{+R}}}
\newcommand{\phml}{\varphi_{{}_{-L}}}
\newcommand{\phmr}{\varphi_{{}_{-R}}}
\newcommand{\phpml}{\varphi_{{}_{\pm L}}}
\newcommand{\phpmr}{\varphi_{{}_{\pm R}}}
\newcommand{\Ei}{\rm Ei}

\newtheorem{theo}{Theorem}[section]
\newtheorem{coro}[theo]{Corollary}
\newtheorem{prop}[theo]{Proposition}
\newtheorem{defi}[theo]{Definition}
\newtheorem{conj}[theo]{Conjecture}
\newtheorem{lem}[theo]{Lemma}
\newcommand{\prf}{\underline{\it Proof.}\ }
\newcommand{\finprf}{\null \hfill {\rule{5pt}{5pt}}\\[2.1ex]\indent}

\pagestyle{empty}
\rightline{November 2006}

\vfill

\begin{center}
{\Large\bf Bosonization and Scale Invariance\\ on Quantum Wires}
\\[2.1em]

\bigskip

{\large
B. Bellazzini$^{a}$\footnote{b.bellazzini@sns.it},  
M. Mintchev$^{b}$\footnote{mintchev@df.unipi.it} 
and P. Sorba$^{c}$\footnote{sorba@lapp.in2p3.fr}}\\

\null

\noindent 

{\it 
$^a$ INFN and Scuola Normale Superiore, Piazza dei Cavalieri 7, 
56127 Pisa,  Italy\\[2.1ex]
$^b$ INFN and Dipartimento di Fisica, Universit\`a di
Pisa, Largo Pontecorvo 3, 56127 Pisa, Italy\\[2.1ex] 
$^c$ LAPTH, 9, Chemin de Bellevue, BP 110, F-74941 Annecy-le-Vieux 
cedex, France}
\vfill

\end{center}

\begin{abstract} 

We develop a systematic approach to bosonization and vertex algebras 
on quantum wires of the form of star graphs. The related bosonic fields 
propagate freely in the bulk of the graph, but interact with its vertex. Our framework
covers all possible interactions preserving unitarity. Special attention is devoted to
the scale invariant interactions, which determine the critical properties of the system. 
Using the associated scattering matrices, we give a complete classification of 
the critical points on a star graph with any number of edges. Critical points 
where the system is not invariant under wire permutations are discovered. 
By means of an appropriate vertex algebra we perform the bosonization of 
fermions and solve the massless Thirring model. In this context we derive an explicit
expression for the conductance and investigate its behavior 
at the critical points. A simple relation between the conductance and the
Casimir energy density is pointed out.

\end{abstract}
\bigskip 
\medskip 
\bigskip

\vfill
\rightline{IFUP-TH 25/2006}
\rightline{LAPTH-1166/06}
\rightline{\tt hep-th/0611090}
\newpage
\pagestyle{plain}
\setcounter{page}{1}

\sect{Introduction} 
\bigskip

Quantum graphs \cite{E}-\cite{KSch} are networks of one-dimensional wires 
connected at nodes. For the first time such structures have been applied some decades 
ago for describing the electron transport in organic molecules. More recently, 
quantum graphs (wires) appeared in the study of interacting one-dimensional 
electron gas \cite{KF}-\cite{Oshikawa:2005fh}. Bosonization represents in this context a basic tool 
for the investigation of various phenomena like transmission through barriers, 
resonant multilead point-contact tunneling and conductance. Motivated by this fact, 
we develop in this paper a general framework for the construction 
of vertex operators and algebras on quantum wires, 
paying special attention on scale invariance.  
Our approach combines results from the spectral theory 
of the Schr\"odinger operator on quantum graphs \cite{Kostrykin:1998gz}-\cite{KSch} 
with an algebraic technique \cite{Liguori:1996xr}-\cite{Mintchev:2005rz} for dealing 
with quantum fields with defects 
(impurities). This combination is quite natural because the junctions 
of the quantum wires can be represented as point-like defects. 
In this context we establish the behavior under scale transformations and classify the 
critical points relative to all dissipationless point-like interactions of a scalar field 
at the junction. Apart from isolated critical points, we discover 
multi-parameter families of such points which are asymmetric  
under permutations of single wires. 
Our framework applies also away from criticality, giving the possibility to analyze 
the renormalization group flows which interpolate between different critical points. 
By coupling the system to an external electric field we show that certain critical points
exhibit exotic conductance properties. In some cases we detect an enhancement 
resulting from an analogue of Andreev's reflection \cite{A} at the junction of the graph. In other
cases  the conductance is depressed with respect to the one of a single wire. Remarkably 
enough, these two regimes correspond to repulsive and attractive 
Casimir forces respectively. Finally, we investigate the effect of 
the four fermion interaction in the bulk solving the massless Thirring model, 
which is a sort of ``relativistic" version of the Tomonaga-Luttinger model.

The paper is organized as follows. In section 2 we present the framework. 
We construct the scalar field $\ph$ and its dual $\phd$ on 
a quantum graph and discuss the Kirchhoff rule associated with some conserved 
currents. The duality transformation $\ph \leftrightarrow \phd$ is also investigated.  
The interaction of $\{\ph,\, \phd\}$ at a junction of a quantum graph 
is studied in section 3, where the scale invariant $S$-matrices are classified. 
The vertex algebra generated by $\{\ph,\, \phd\}$ 
and the statistics of the vertex operators are described in section 4. 
Using the vertex algebra, we perform in section 5 
the bosonization of fermions, derive the corresponding 
correlation functions and compute the conductance. Here we discuss 
also the relation between conductance and Casimir effect. In section 6 we introduce 
non-trivial bulk interactions and analyze the massless Thirring model on a 
star graph. The impact of the bulk interaction on the conductance is also 
determined. Section 7 makes contact with some previous work on quantum wires and 
boundary conformal field theory. It contains also our conclusions and some 
ideas for further developments.

\bigskip 

\sect{The framework} 

{}For simplicity we consider in this paper quantum wires of 
the form of a star graph\footnote{Such graphs represent 
the building blocks for general graphs and are essential in 
experiments \cite{Terr} as well.} $\Gamma$ shown in Fig. 1. 
\vskip 0.5truecm 
\setlength{\unitlength}{1,25mm}
\begin{picture}(20,20)(-25,20) 
\put(25.2,0.7){\makebox(20,20)[t]{$\bullet$}}
\put(28.5,1){\makebox(20,20)[t]{$V$}}
\put(42,11){\makebox(18,22)[t]{$E_1$}}
\put(33,17){\makebox(20,20)[t]{$E_2$}}
\put(9,3.5){\makebox(20,20)[t]{$E_i$}}
\put(34.5,-12){\makebox(20,20)[t]{$E_n$}}
\thicklines 
\put(35,20){\line(1,1){16}}
\put(35,20){\line(-1,0){19}}
\put(35,20){\line(1,-1){13}}
\put(35,20){\line(1,3){6}}
\put(20,3){\makebox(20,20)[t]{$.$}}
\put(20.9,5){\makebox(20,20)[t]{$.$}}
\put(23.8,6.6){\makebox(20,20)[t]{$.$}}
\put(20,-3){\makebox(20,20)[t]{$.$}}
\put(20.9,-5){\makebox(20,20)[t]{$.$}}
\put(23.8,-6.6){\makebox(20,20)[t]{$.$}}
\put(46,31){\vector(1,1){0}}
\put(46,9){\vector(1,-1){0}}
\put(40,35){\vector(1,3){0}}
\put(20,20){\vector(-1,0){0}}
\end{picture}
\vskip 2.5 truecm
\centerline{Figure 1: A star graph $\Gamma$ with $n$ edges.}
\bigskip 
\noindent 
Each point $P$ in the bulk $\Gamma \setminus V$ of the graph $\Gamma$ 
belongs to some edge $E_i$ and can be parametrized by the pair $(x,i)$, 
where $x>0$ is the distance of $P$ from the vertex (junction) $V$ along $E_i$. 
The embedding of $\Gamma$ and the relative position of 
its edges in the ambient space are irrelevant in what follows. 

\subsection{The scalar field $\ph$ and its dual $\phd$}

The basic ingredient for bosonization is the massless scalar 
field $\ph$, which satisfies 
\be
\left (\prt_t^2 - \prt_x^2 \right )\ph (t,x,i) = 0\, , 
\qquad x> 0 \, ,  \;  i=1,...,n \, , 
\label{eqm1}
\ee 
in the bulk of $\Gamma$ and the vertex boundary condition 
\be 
\sum_{j=1}^n \left [A_i^j\, \ph (t,0,j) + B_i^j (\prt_x\ph ) (t,0,j)\right ] = 0\, , 
\qquad \forall \, t\in \RR\, , \; i=1,...,n\, , 
\label{bc} 
\ee 
$A$ and $B$ being in general two $n\times n$ complex matrices. Clearly the pairs 
$\{A,\, B\}$ and $\{CA,\, CB\}$, where $C$ is any invertible matrix, 
define equivalent boundary conditions. 

The results of \cite{Kostrykin:1998gz} imply that the Hamiltonian of the system is self-adjoint, 
provided that\footnote{We denote by ${}^*$ the Hermitian conjugation.} 
\be 
A\, B^* - B\, A^* = 0 \,  ,
\label{cond0} 
\ee
and the composite matrix $(A,B)$ has rank $n$. In what follows we refer 
to the latter as the rank condition and stress that the matrices $A$ and $B$ parametrize 
{\it all possible} self-adjoint extensions of the Hamiltonian from $\Gamma \setminus V$ to 
$\Gamma $. Requiring that $\ph$ is real (Hermitian) and 
imposing invariance under time reversal\footnote{In other words the existence of a 
{\it antiunitary} operator $T$ such that $T\ph(t,x)T^{-1} = \ph(-t,x)$.}, 
one infers from (\ref{bc}) that there exist an invertible matrix $C$, such that 
\be 
{\overline A} = C\, A\, , \qquad  {\overline B} = C\, B\, , 
\label{real}
\ee
where the bar stands for complex conjugation. From (\ref{real}) and 
the rank condition it follows that $C\, {\overline C} ={\overline C}\, C=\II_n$, where 
$\II_n$ is the $n\times n$ identity matrix. Thus, 
setting $C_\theta = (\e^{\ri \theta}\II_n + \e^{-\ri \theta} C)$, 
we conclude that $\{C_\theta A,\, C_\theta B\}$ are real matrices for any $\theta \in \RR$. 
Since $C_\theta$ is invertible if $-\e^{2\ri \theta}$ is not an eigenvalue of $C$, 
the boundary conditions defined by $\{A,\, B\}$ and $\{C_\theta A,\, C_\theta B\}$ are equivalent. 
Therefore, without loss of generality one can assume in what follows that $A$ and $B$ are 
{\it real}. Accordingly, (\ref{cond0}) takes the form 
\be 
A\, B^t - B\, A^t = 0 \,  ,
\label{cond1} 
\ee 
where the apex $t$ denotes transposition. 

{}For invertible $B$ eqs. (\ref{eqm1}, \ref{bc}) define a variational problem with the action 
\be 
I [\ph] = I_0[\ph] 
+ \frac{1}{2} \sum_{i,j=1}^n \int_{-\infty}^\infty \rd t\, \ph(t,0,i) \left (B^{-1}A\right )_i^j  \ph(t,0,j) \, . 
\label{action}
\ee 
where 
\be
I_0[\ph] = \frac{1}{2} \sum_{i=1}^n \int_{-\infty}^\infty \rd t \int_0^\infty \rd x \left [(\prt_t \ph ) (\prt_t \ph ) - 
(\prt_x \ph ) (\prt_x \ph )\right ](t,x,i) \, . 
\label{freeaction}
\ee  
The dual field $\phd$ is defined by the relations 
\be 
\prt_t \phd (t,x,i) = - \prt_x \ph (t,x,i)\, , \quad  
\prt_x \phd (t,x,i) = - \prt_t \ph (t,x,i)\, ,
\qquad x> 0 \, ,  \;  i=1,...,n \, ,  
\label{eqm2}
\ee 
which imply that 
\be
\left (\prt_t^2 - \prt_x^2 \right )\phd (t,x,i)= 0\, , 
\qquad x> 0 \, ,  \;  i=1,...,n  
\label{eqm3}
\ee
as well. 

The problem of quantizing (\ref{eqm1}, \ref{eqm2}, \ref{eqm3}) with the 
boundary condition (\ref{bc}) and initial conditions fixed by 
the equal-time canonical commutation relations 
\be
[\ph (0,x_1,i_1)\, ,\, \ph (0,x_2,i_2)] = 
[\phd (0,x_1,i_1)\, ,\, \phd (0,x_2,i_2)] = 0\, , \qquad 
\label{initial1}
\ee
\be 
[(\prt_t\ph )(0,x_1,i_1)\, ,\, \ph (0,x_2,i_2)] = 
[(\prt_t\phd )(0,x_1,i_1)\, ,\, \phd (0,x_2,i_2)] =
-\ri \delta_{i_1}^{i_2}\, \delta (x_1-x_2) \, .
\label{initial2}
\ee 
has a unique solution. It can be written in the form \cite{Bellazzini:2006jb} 
\be
\ph (t,x,i) = \int_{-\infty}^{\infty} \frac{\rd k}{2\pi \sqrt
{2|k|}}
\left[a^{\ast i}(k) \e^{\ri (|k|t-kx)} +
a_i (k) \e^{-\ri (|k|t-kx)}\right ] \,  , 
\label{sol1}
\ee
\be
\phd (t,x,i) = \int_{-\infty}^{\infty} 
\frac{\rd k\, \varepsilon (k)}{2\pi \sqrt {2|k|}} 
\left[a^{\ast i}(k) \e^{\ri (|k|t-kx)} +
a_i (k) \e^{-\ri (|k|t-kx)}\right ] \,  ,  
\label{sol2}
\ee 
where $\varepsilon (k)$ is the sign function and 
$\{a_i(k),\, a^{* i}(k)\, :\, k\in \RR\}$ generate the boundary 
(reflection-transmission) algebra \cite{Liguori:1996xr}-\cite{Mintchev:2005rz} 
corresponding to the boundary condition (\ref{bc}). This is an associative algebra $\A$ 
with identity element $\bf 1$, whose generators $\{a_i(k),\, a^{* i}(k)\, :\, k\in \RR\}$ 
satisfy the commutation relations  
\bea
&a_{i_1}(k_1)\, a_{i_2}(k_2) -  a_{i_2}(k_2)\, a_{i_1}(k_1) = 0\,  ,
\label{ccr1} \\
&a^{\ast i_1}(k_1)\, a^{\ast i_2}(k_2) - a^{\ast i_2}(k_2)\,
a^{\ast i_1}(k_1) = 0\,  ,
\label{ccr2} \\
&a_{i_1}(k_1)\, a^{\ast i_2}(k_2) - a^{\ast i_2}(k_2)\,
a_{i_1}(k_1) = 
2\pi \left [\delta_{i_1}^{i_2} \delta(k_1-k_2) +
S_{i_1}^{i_2}(k_1) \delta(k_1+k_2)\right ] {\bf 1}\,  , 
\nonumber \\ 
\label{ccr3}
\eea 
and the constraints\footnote{Summation over repeated upper and lower indices is 
understood throughout the paper.}
\be
a_i(k) = S_i^j (k) a_j (-k) \, , \qquad 
a^{\ast i}(k) = a^{\ast j}(-k) S_j^i (-k)\, .    
\label{constr1}
\ee 
The $S$-matrix in (\ref{ccr3}, \ref{constr1}) equals \cite{Kostrykin:1998gz} 
\be 
S(k) = -(A+\ri kB )^{-1} (A-\ri kB) 
\label{S1}
\ee
and has the following simple physical interpretation. In spite of the fact that 
$\{\ph\, , \phd\}$ propagate freely in the bulk of the graph, they interact 
with a specific external potential localized in the vertex and codified by 
the boundary condition (\ref{bc}). 
This interaction leads to non-trivial reflection and transmission. 
The associated $S$-matrix is given by (\ref{S1}): the diagonal 
element $S_i^i(k)$ is the reflection amplitude on the edge $E_i$, 
whereas $S_i^j(k)$ with $i\not=j$ is the transmission amplitude 
from $E_i$ to $E_j$. Since the vertex $V$ can be viewed as a sort of impurity, 
it is not surprising that the algebra $\A$, appearing in the 
context of quantum field theory with boundaries 
or defects \cite{Liguori:1996xr}-\cite{Mintchev:2005rz}, represents a convenient tool 
for the analysis of $\{\ph ,\phd\}$ on the star graph $\Gamma$. In fact, $\A$ provides a simple 
algebraic description of all self-adjoint extensions of the bulk Hamiltonian to the whole graph. 

We assume in what follows that $A$ and $B$ are such that 
\be
\int_{-\infty}^{\infty} \frac{\rd k}{2\pi } \e^{\ri kx} S_i^j(k) = 0\, . 
\qquad x>0\, , 
\label{compl1}
\ee 
This condition implies the absence of bound states, ensuring that 
$\A$ is a complete basis \cite{Bellazzini:2006jb} in agreement with (\ref{sol1}, \ref{sol2}). 
Let us mention also that (\ref{S1}) is unitary 
\be 
S(k)^*=S(k)^{-1} \, , 
\label{unit1}
\ee 
and satisfies Hermitian analyticity 
\be 
S(k)^*=S(-k)\, .  
\label{Ha}
\ee 
Because of (\ref{real}), one has actually that $S$ is symmetric
\be 
S(k)^t=S(k)\, ,  
\label{symm}
\ee
which is equivalent to invariance under time reversal. 
Combining (\ref{unit1}) and (\ref{Ha}) one gets 
\be 
S(k)\, S(-k) = \II_n \, , 
\label{unit2} 
\ee
which ensures the consistency of the constraints (\ref{constr1}). 
As it should be expected, the boundary conditions associated with 
$\{A,B\}$ and $\{CA,CB\}$ with invertible $C$, lead to the same $S$.

\subsection{Symmetries and Kirchhoff's rule}

The concept of symmetry on quantum graphs needs special attention. Let 
$j_\nu (t,x,i)$ be a conserved current, i.e. 
\be 
\prt_t j_t(t,x,i) - \prt_x j_x(t,x,i)=0 \, . 
\label{curr0}
\ee
The time derivative of the corresponding charge is  
\be 
\prt_t \sum_{i=1}^n \int_0^\infty \rd x\, j_t(t,x,i) = 
\sum_{i=1}^n \int_0^\infty \rd x\, \prt_x j_x(t,x,i) = \sum_{i=1}^n  j_x(t,0,i)\, ,
\label{charge0}
\ee 
implying charge conservation if and only if Kirchhoff's rule  
\be 
\sum_{i=1}^n  j_x(t,0,i) = 0 
\label{kirgen}
\ee 
holds in the vertex $V$. This fact is essential in our context. 
The invariance of the equations of motion (\ref{eqm1},\ref{eqm2}) 
under time translations implies the conservation of the current 
\be 
\theta_{tt}(t,x,i) = \frac{1}{2}: \left [ (\prt_t \ph )(\prt_t \ph ) - 
\ph (\prt_x^2 \ph ) \right ]:(t,x,i)\, , 
\ee
\be
\theta_{tx}(t,x,i) = \frac{1}{2} :\left [ (\prt_t \ph)( \prt_x \ph) - 
\ph (\prt_t \prt_x \ph )\right ] :(t,x,i)\, , 
\label{emt}
\ee
where $:\cdots :$ denotes the normal product in the algebra $\A$. The associated Kirchhoff rule 
\be 
\sum_{i=1}^n  \theta_{tx}(t,0,i) = 0 
\label{kirteta}
\ee 
is satisfied by construction, being a consequence \cite{Kostrykin:1998gz} of 
(\ref{bc}), (\ref{cond1}) and the rank condition on $A$ and $B$. Eq. (\ref{kirteta}) 
guarantees energy conservation (no dissipation) and represents a meeting point 
between boundary conformal field theory \cite{Cardy:1984bb}-\cite{Cardy:2004hm} 
and the concept of scale invariance on a star graph with $n\geq 2$ edges. 
We will elaborate on this point in section 7, keeping for the moment 
the discussion as general as possible. 
 
The equations of motion (\ref{eqm1},\ref{eqm2}) are also invariant under 
the transformations 
\be 
\ph (t,x,i) \longmapsto \ph (t,x,i) + c\, , \qquad 
\phd (t,x,i) \longmapsto \phd (t,x,i) + {\widetilde c}\, , \quad c,\, {\widetilde c} \in \RR \, ,
\label{shiftsymm}
\ee
which implies the conservation of the currents 
\be 
k_\nu (t,x,i) = \prt_\nu \ph (t,x,i)\, , \qquad 
\kt_\nu (t,x,i)= \prt_\nu \phd (t,x,i)\, , \qquad \nu = t,\, x\, . 
\label{curr1}
\ee
Using the solution (\ref{sol1}) and the constraints (\ref{constr1}) one finds that $k_\nu$ also satisfies 
a Kirchhoff's rule 
\be 
\sum_{i=1}^n k_x(t,0,i) = 0 \, , 
\label{kir1}
\ee 
provided that 
\be 
\sum_{j=1}^n S_i^j(k) = 1 \, ,\qquad \forall \; i=1,...,n\, , \; k\in \RR \, .  
\label{kir2}
\ee 
Analogously, 
\be 
\sum_{i=1}^n \kt_x(t,0,i) = 0 \, , 
\label{kir3}
\ee 
holds if 
\be 
\sum_{j=1}^n S_i^j(k) = -1 \, ,\qquad \forall \; i=1,...,n\, , \; k\in \RR \, .  
\label{kir4}
\ee 
Combining (\ref{kir2}) and (\ref{kir4}) one draws the important conclusion that 
Kirchhoff's rule cannot be satisfied simultaneously for both $k_\nu$ and its dual $\kt_\nu$. 
Accordingly, at most one of the charges 
\be 
Q = \sum_{i=1}^n \int_0^\infty \rd x\, k_t(t,x,i)\, , \qquad 
{\widetilde Q} =  \sum_{i=1}^n \int_0^\infty \rd x\, \kt_t(t,x,i)\, ,
\label{charges}
\ee 
is $t$-independent, which is a first indication that the duality transformation $\ph \leftrightarrow \phd$, 
discussed few lines below, is not a symmetry of the boundary value problem in consideration. 

{}For the classification of the boundary conditions at the junction, performed in the next section, 
it is useful to translate (\ref{kir2}, \ref{kir4}) in terms of the matrices $A$ and $B$ entering 
eq. (\ref{bc}). For this purpose we introduce the vector ${\bf v}=(1,1,...,1)$ and observe that 
\be 
(\ref{kir2}) \Longleftrightarrow 
{\bf v} \in {\rm Ker} \left [S(k)-\II_n\right ] = {\rm Ker}\, A\, , 
\label{kir5}
\ee
where the equality between the two kernels follows 
(see Lemma 3.17 of \cite{KSch}) from the explicit form (\ref{S1}) of $S$. 
In simple words, Kirchhoff's rule (\ref{kir1}) is satisfied if and only if 
the entries along each line of the matrix $A$ sum up to 0. 
In a similar way one can deduce that 
\be  
(\ref{kir4}) \Longleftrightarrow 
{\bf v} \in {\rm Ker} \left [S(k)+\II_n\right ] 
= {\rm Ker}\, B \, .  
\label{kir6}
\ee 

We observe in conclusion that the equations of motion (\ref{eqm1},\ref{eqm2}) and the 
initial conditions (\ref{initial1}, \ref{initial2}) 
are invariant under the exchange $\ph \leftrightarrow \phd$, 
but the boundary condition (\ref{bc}) breaks down the 
symmetry.  This fact suggests to consider 
a similar but nonequivalent boundary value problem, defined by replacing (\ref{bc}) with 
\be 
\sum_{j=1}^n \At_i^j\, \phd (t,0,j) + \Bt_i^j (\prt_x\phd\, ) (t,0,j) = 0\, , 
\qquad \forall \, t\in \RR\, , \; i=1,...,n\, , 
\label{bc1} 
\ee
where $\At$ and $\Bt$ satisfy the same conditions as $A$ and $B$. 
The solution of the new problem is still given by (\ref{sol1},\ref{sol2}) with the substitution 
\be 
S(k) \mapsto \St (k) = (\At+\ri k\Bt )^{-1} (\At-\ri k\Bt)  
\label{S2}
\ee 
in the algebra $\A$. Summarizing, the duality transformation $\ph \leftrightarrow \phd$ 
relates the problem defined by (\ref{bc}) to the one defined by (\ref{bc1}). 
 
\bigskip

\sect{Interaction at the junction} 

As already observed, the fields $\{\ph ,\phd\}$ freely propagate in the 
bulk of the star graph, but interact at its vertex. The most general interaction, 
preserving unitarity, is fixed by the boundary condition (\ref{bc}), where 
$A$ and $B$ satisfy the conditions given in the previous section. For each $n\geq 1$
the admissible pairs $\{A,B\}$ define a family $\cS_n$ of $S$-matrices (\ref{S1}) with 
intriguing structure. Before discussing the general features of $\cS_n$, it is 
instructive to give some examples. 

\subsection{Examples}

We start with the familiar Dirichlet 
\be 
\ph (t,0,1)=\ph (t,0,2)=\dots =\ph (t,0,n) = 0 
\label{dir1}
\ee 
and Newmann 
\be 
(\prt_x\ph) (t,0,1)=(\prt_x\ph) (t,0,2)=\dots =(\prt_x\ph) (t,0,n)=0  
\label{new1}
\ee 
boundary conditions, corresponding to $A=1,\,  B=0$ and $A=0,\,  B=1$ respectively. 
The associated $S$-matrices are $S_{\rm D} = -\II_n$ and $S_{\rm N} = \II_n$. For 
$n\geq 2$ there are many possible generalizations of the mixed (Robin) 
boundary condition. Let us consider for instance 
\be 
\ph (t,0,1)=\ph (t,0,2)=\dots =\ph (t,0,n)\, , \qquad 
\sum_{i=1}^n (\prt_x\ph) (t,0,i) = \eta \ph (t,0,n)\, .  
\label{rob1}
\ee 
In this case 
\be
A=\left(\begin{array}{cccccc}
1&-1&0& \cdots &0&0\\ 
0&1&-1& \cdots &0&0\\
\vdots&\vdots&\vdots&\cdots&\vdots&\vdots\\
0&0&0&\cdots&1&-1\\
0&0&0&\cdots&0&-\eta 
\end{array}\right)\, , \qquad 
B=\left(\begin{array}{cccccc}
0&0&0& \cdots &0&0\\ 
0&0&0& \cdots &0&0\\
\vdots&\vdots&\vdots&\cdots&\vdots&\vdots\\
0&0&0&\cdots&0&0\\
1&1&1&\cdots&1&1 
\end{array}\right)\, , 
\label{rob2}
\ee
leading to 
\be 
S(k) = \frac{1}{nk+\ri \eta}
\left(\begin{array}{ccccc}
(2-n)k-\ri \eta&2k&2k& \cdots &2k\\ 
2k&(2-n)k-\ri \eta&2k& \cdots &2k\\
\vdots&\vdots&\vdots&\cdots&\vdots\\ 
2k&2k&2k&\cdots&(2-n)k-\ri \eta  
\end{array}\right)\, .
\label{rob3}
\ee 

The boundary conditions (\ref{dir1},\ref{new1},\ref{rob1}) 
are symmetric under edge permutations\footnote{A systematic discussion 
of this kind of boundary conditions is given in \cite{E1}.}, 
which is clearly not the case in general. 
A simple asymmetric example is defined by 
\be 
A =\frac{2}{3\rho} 
\left(\begin{array}{ccc}
-1&1&0\\ 
0&-1&1\\
0&0&0  
\end{array}\right)\, ,  \qquad 
B =
\left(\begin{array}{ccc}
1&0&0\\ 
0&0&0\\
1&1&1  
\end{array}\right)\, , 
\qquad  \rho >0  \, . 
\label{ex1}
\ee 
The associated $S$-matrix reads 
\be 
S (k) = \frac{1}{3(1-\ri \rho k)} 
\left(\begin{array}{ccc}
-1-3\ri \rho k&2 &2 \\ 
2&-1&2-3\ri \rho k\\
2 &2-3\ri \rho k&-1 
\end{array}\right)\, , 
\label{ex2}
\ee
which is not invariant under the permutations $1\leftrightarrow 2$ and 
$1\leftrightarrow 3$. 

\subsection{Duality and scale invariance}

The basic concepts for analyzing the general structure of $\cS_n$ are 
{\it duality} and {\it scale invariance}, 
combined with Kirchhoff's rules (\ref{kir1},\ref{kir3}). 
Let us describe first the duality in $\cS_n$. 
It is easily seen that the transformation \cite{Kostrykin:1998gz} 
\be 
\{A,B\} \longmapsto \{-B,A\} 
\label{sd1}
\ee 
defines an admissible boundary condition and induces the mapping 
\be 
S(k) \longmapsto  -S (k^{-1}) 
\label{sd2}
\ee 
on $\cS_n$. Therefore (\ref{sd1}) relates high and low momenta, justifying 
the term duality transformation in $\cS_n$. It is worth mentioning that 
if $\{A,B\}$ implies Kirchhoff's rule for the current $k_\nu$, then 
its dual image $\{-B,A\}$ enforces Kirchhoff's rule for 
the dual current $\kt_\nu$. We observe also that (\ref{sd1}) does not preserve in general 
the completeness condition (\ref{compl1}). In fact, it follows from 
(\ref{sd2}) that resonant states are dual images of bound states and vice 
versa\footnote{See for example (\ref{ex2})}. 

Scale invariance determines the critical points and plays therefore 
a distinguished role. A simple example of a scale invariant $S$-matrix 
is obtained by setting $\eta=0$ in (\ref{rob1}), which leads to 
\be  
S = \frac{1}{n}
\left(\begin{array}{ccccc}
(2-n)&2&2& \cdots &2\\ 
2&(2-n)&2& \cdots &2\\
\vdots&\vdots&\vdots&\cdots&\vdots\\ 
2&2&2&\cdots&(2-n)   
\end{array}\right)\, .
\label{sc1}
\ee 
The $k$-independence of (\ref{sc1}) is actually a general feature of any 
scale invariant $S$-matrix and has two direct implications: 

(i) the condition (\ref{compl1}) holds, implying the absence of bound states; 

(ii) the duality transformation (\ref{sd2}) maps any scale 
invariant matrix $S$ in $-S$ and in terms of fields is realized by the 
mapping $\ph \leftrightarrow \phd$. 

In the scale invariant case the matrices $A$ and $B$ have a particular form. 
Since the interplay between the two terms in the boundary condition (\ref{bc}) 
involves a dimensional parameter, it is clear that these terms must decouple  
at a critical point. This is only possible if, up to a multiplication with 
a common invertible matrix, the non-vanishing lines 
of $A$ are complementary to those of $B$ (see e.g. (\ref{rob2}) for 
$\eta =0$). In other words, if $A$ has $0\leq p\leq n$
non-vanishing lines at certain position, $B$ has $n-p$ such lines in the complementary
position. Combining this observation with the rank condition we deduce that at any 
critical point ${\rm rank}_A = p$ and ${\rm rank}_B = n-p$. 
One can use therefore the parameter $p$ for the classification 
of the scale invariant $S$-matrices. The limiting cases $p=0$ and $p=n$ 
are simple: one has $A=0$ and $B=0$ respectively, leading to 
$S_{\rm N}=\II_n$ and $S_{\rm D}=-\II_n$. So, from now on we 
concentrate on the range $0< p < n$.

\subsection{Critical points for $n=2,3$}

In what follows we focus on $S$-matrices satisfying Kirchhoff's 
rule (\ref{kir1}). For $n=2$ we have only the case $p=1$ associated with 
\be 
A =
\left(\begin{array}{cc}
a_{11}&-a_{11}\\ 
0&0
\end{array}\right)\, , \qquad 
B =
\left(\begin{array}{cc}
0&0\\ 
b_{21}&b_{22}
\end{array}\right)\, ,
\label{nsc}
\ee 
where the entries along the lines of $A$ sum to up to $0$ as required by 
Kirchhoff's rule. Inserting (\ref{nsc}) in (\ref{S1}) and imposing 
(\ref{cond1}) and the rank condition one gets
\be 
S =
\left(\begin{array}{cc}
0&1\\ 
1&0
\end{array}\right)\, . 
\label{sc}
\ee 
This $S$-matrix describes complete transmission without reflection 
and produces actually the free field on $\RR = E_1\cup E_2$. 
Relaxing  Kirchhoff's rule (\ref{kir1}), one has the more interesting 
one-parameter family of scale invariant $S$-matrices 
\be
S = \frac{1}{1+\alpha^2}
\left(\begin{array}{cc}
\alpha^2-1&-2\alpha\\ 
-2\alpha&1-\alpha^2
\end{array}\right)\, , \qquad \alpha \in \RR\, , 
\label{scnk}
\ee 
which have been studied in \cite{Mintchev:2005rz, Bachas:2001vj}. 

The case $n=3$ has a richer structure. For $p=2$ one finds 
\be 
S =\frac{1}{3} 
\left(\begin{array}{ccc}
-1&2&2\\ 
2&-1&2\\
2&2&-1  
\end{array}\right)\, , 
\label{sc2}
\ee 
which coincides with (\ref{sc1}) for $n=3$ and represents the scattering matrix 
for all boundary conditions given 
(up to a multiplication with a common invertible matrix) by 
\be 
A =  
\left(\begin{array}{ccc}
a_{11}&a_{12}&-a_{11}-a_{12}\\ 
a_{21}&a_{22}&-a_{21}-a_{22}\\
0&0&0  
\end{array}\right)\, , \qquad  
B =
\left(\begin{array}{ccc}
0&0&0\\ 
0&0&0\\
b_{31}&b_{32}&b_{33}  
\end{array}\right) \, , 
\label{sc3}
\ee 
The critical point (\ref{sc2}) is invariant under edge 
permutations and has been discovered in \cite{NFLL} by studying the
renormalization group flow of a specific model of resonant point-contact 
tunnelling in a system with three wires. 

In the case $p=1$ one finds a one-parameter family of critical points 
\be 
S =\frac{1}{1+\alpha +\alpha^2} 
\left(\begin{array}{ccc}
\alpha +1&-\alpha&\alpha(\alpha +1)\\ 
-\alpha&\alpha(\alpha +1)&\alpha +1\\
\alpha(\alpha +1)&\alpha +1&-\alpha  
\end{array}\right)\, , \qquad \alpha \in \RR\, , 
\label{sc4}
\ee 
corresponding to the class of boundary conditions 
\be 
A =  
\left(\begin{array}{ccc}
a_{11}&a_{12}&-a_{11}-a_{12}\\ 
0&0&0\\
0&0&0  
\end{array}\right)\, , \qquad  
B =
\left(\begin{array}{ccc}
0&0&0\\ 
b_{21}&b_{22}&b_{23}\\
b_{31}&b_{32}&b_{33}  
\end{array}\right) \, , 
\label{sc5}
\ee 
The rank condition implies that at least one of the parameters 
$a_{11}$ and $a_{12}$ does not vanish. Renumbering the edges 
and rescaling $A$, one can take without loss of generality $a_{12}=1$. Then 
$\alpha = a_{11}$. Differently from (\ref{sc2}), there are no 
edge permutations leaving invariant (\ref{sc4}) for generic $\alpha$. 
To our knowledge the family of critical points (\ref{sc4}) is new and 
has not been previously investigated. The peculiar behavior of the 
conductance in this case is described in section 5.3. 

The critical points (\ref{sc}, \ref{sc2}, \ref{sc4}) are mapped by the duality
transformation (\ref{sd1}) to critical points, where the Kirchhoff rule
(\ref{kir3}) for the dual current $\kt_\nu$ holds.

\subsection{Critical points for $n\geq 4$}

The explicit results for $n=3$ suggest that besides isolated critical points, 
the phase diagram for $n\geq 4$ involves also multi-parameter families of such points. 
We would like now to determine the number of these parameters and clarify their meaning. For 
this purpose we first observe that the scale invariant $S$-matrices can be written 
in the form \cite{KSch} 
\be 
S=\II_n - 2P_{{\rm Ker} B} = -\II_n + 2P_{{\rm Ker} A}\, , 
\label{kerrep}
\ee 
where $P$ is the projection operator. Using that the non-vanishing lines of $A$ are 
complementary to that of $B$, from (\ref{cond1}) we deduce that in the scale invariant case 
\be 
A\, B^t = 0\, ,\qquad B\, A^t =0 \, . 
\label{critcond1}
\ee 
Combined with the rank condition, eqs. (\ref{critcond1}) 
imply that the lines of the matrix $B$ form 
a basis in ${\rm Ker} A$ and vice versa the lines of $A$ provide a basis in ${\rm Ker} B$. 
One has in addition that ${\rm Ker} A$ and 
${\rm Ker} B$ are orthogonal and, assuming Kirchhoff's rule (\ref{kir1}), that the 
vector ${\bf v}=(1,1,...,1)$ is orthogonal to ${\rm Ker} B$. Therefore, 
${\rm Ker} B$ is a $p$-dimensional subspace embedded in $\RR^{(n-1)}$. 

Suppose we take now a $n\times n$-matrix $A$ with $0< p< n$ non-vanishing 
lines. Our goal is to determine the number $N(n,p)$ of parameters involved in the corresponding 
$S$-matrix. According to our previous discussion, $N(n,p)$ counts the parameters 
needed in order to fix uniquely the position of ${\rm Ker} B$ in $\RR^{(n-1)}$. One has 
\be 
N(n,p) = p(n-1-p)\, , 
\label{number}
\ee 
which can be easily derived as follows. 
Let $\{{\bf a}_1,...,{\bf a}_p\}$ be a basis in ${\rm Ker} B$ and 
let the vectors $\{{\bf b}_1,...,{\bf b}_{n-1-p}\}$ complement it to a basis in $\RR^{(n-1)}$. 
The position of ${\rm Ker} B$ in $\RR^{(n-1)}$ is 
determined by fixing all possible scalar products of the type 
${\bf a}_i\cdot {\bf b}_j$, whose number is precisely (\ref{number}). 

Summarizing, the number of parameters 
characterizing any family of critical points has a simple geometric 
meaning related to the embedding of 
${\rm Ker} B$ in $\RR^{(n-1)}$. It is also clear that a 
similar argument, using ${\rm Ker} A$ instead of 
${\rm Ker} B$, gives the same result. Solving $N(n,p)=0$, one finds $p=n-1$ which 
gives the constant solution (\ref{sc1}). In all other cases ($0<p<n-1$) the 
$S$-matrix involves $N(n,p)>0$ real parameters. In spite of this fact, the trace of $S$ 
depends exclusively on $n$ and $p$. In fact, one gets from (\ref{kerrep}) 
\be 
{\rm Tr}\, S = n-2p \, ,   
\label{tr1}
\ee
which will be useful in what follows. 

Let us apply now (\ref{number}) to the case $n=4$. For $p=3$ one finds $S=(\ref{sc1})$ with $n=4$. 
In both of the remaining two cases ($p=1,2$) eq. (\ref{number}) 
predicts a two-parameter family of critical points, 
which is confirmed by the explicit computation. The form of the corresponding $S$ matrices 
is a bit involved and is reported in the appendix. 

We observe finally that $S$ depends on $p(n-p)$ real parameters 
if Kirchhoff's rule (\ref{kir1}) is not imposed.

\subsection{Flows}

Once the critical points have been classified, one can investigate the renormalization 
group flows among them. The renormalization group analysis (based on instanton gas 
expansion and strong-weak coupling duality) performed in \cite{NFLL}, suggests the existence of a flow 
between $S_{\rm N} = \II_3$ and (\ref{sc2}). Our formalism enables one to construct 
such a flow explicitly. Let us consider in fact the boundary conditions defined by 
\be 
A =  \rho
\left(\begin{array}{ccc}
-1&0&1\\ 
0&-1&1\\
0&0&0  
\end{array}\right)\, , \qquad  
B =
\left(\begin{array}{ccc}
1&0&0\\ 
0&1&0\\
1&1&1  
\end{array}\right) \, , \qquad \rho \geq 0 \, . 
\label{flow1}
\ee 
The resulting $S$-matrix is 
\be 
S (k) =\frac{1}{(k+\ri \rho )(k+3\ri \rho )}
\left(\begin{array}{ccc}
k^2+2\ri \rho k+\rho^2&-2\rho^2&2\ri \rho (k+\ri \rho )\\ 
-2\rho^2&k^2+2\ri \rho k+\rho^2&2\ri \rho (k+\ri \rho )\\
2\ri \rho (k+\ri \rho)&2\ri \rho (k+\ri \rho )&k^2-\rho^2  
\end{array}\right)\, , 
\label{flow2}
\ee
which indeed interpolates between $\II_3$ ($\rho=0$) and 
(\ref{sc2}) ($\rho \to \infty$). From (\ref{action}) the action along this flow is 
\bea 
I [\ph] = I_0[\ph ] - \frac{\rho}{2} \int_{-\infty}^\infty \rd t\, [\ph^2(t,0,1) + \ph^2(t,0,2) +2 \ph^2(t,0,3) 
\qquad \quad 
\nonumber \\
-2 \ph (t,0,1)\ph (t,0,3) - 2 \ph (t,0,2)\ph (t,0,3) ]\, . 
\label{flowaction}
\eea 

Another interesting flow is defined by (\ref{ex2}). 
In the limit $\rho \to 0$ one gets the critical point 
(\ref{sc2}), whereas for $\rho \to \infty$ one finds 
\be 
S = 
\left(\begin{array}{ccc}
1&0&0\\ 
0&0&1\\
0&1&0  
\end{array}\right)\, , 
\label{flow3}
\ee
which coincides with the critical point $\alpha = 0$ in the family (\ref{sc4}).  
The flow (\ref{ex2}) therefore interpolates between the isolated critical point 
(\ref{sc2}) and the family (\ref{sc4}). 

Let us observe also that bound states are absent and  
Kirchhoff's rule (\ref{kir1}) is satisfied along both the above flows. 
There exist therefore $S$-matrices which are not scale invariant but 
preserve (\ref{kir1}). 

\subsection{Inverse scattering} 

We discuss now the reconstruction of the boundary condition (\ref{bc}) from the 
scattering matrix. This point is relevant because $S(k)$ is the only 
physical observable in the above framework. Remarkably enough, for recovering 
the matrices $A$ and $B$ one needs only the value 
\be 
S_0 = S(k_0) \, , 
\label{inv1}
\ee 
of the $S$-matrix at arbitrary but fixed momentum $k_0\not=0$. According to 
(\ref{unit1}, \ref{symm}), we have 
\be 
S_0^* = S_0^{-1} \, , \qquad S_0^t = S_0 \, . 
\label{inv2}
\ee 
Now, following \cite{KoS,H2,KSch} we set 
\be 
A = \frac{1}{2}C (\II_n - S_0) \, , \qquad 
B = -\frac{\ri}{2k_0} C (\II_n + S_0) \, ,  
\label{inv3}
\ee 
where $C$ is some invertible matrix to be fixed in a moment. One gets from (\ref{inv3}) 
\be 
A B^t = \frac{\ri }{2k_0}C (\II_n-S_0^2)C^t \, , \qquad 
A + \ri k_0 B = C \, , 
\label{inv4}
\ee 
which imply (\ref{cond1}) and the rank condition respectively. Finally, 
we choose $C$ in such a way that $A$ and $B$ are real. Using (\ref{inv2}) we 
see that one can take for this purpose 
\be 
C= \e^{\ri \theta}\II_n -  \e^{-\ri \theta} S_0^* \, , 
\label{inv5}
\ee 
{}for any $\theta$ such that $\e^{-2\ri \theta}$ is not among the eigenvalues of $S_0$. 
Plugging (\ref{inv3}) in (\ref{S1}), one gets 
\be 
S(k)=[(k+k_0)\II_n + (k-k_0)S_0]^{-1} [(k-k_0)\II_n + (k+k_0)S_0]\, , 
\label{inv6}
\ee
which obviously satisfies (\ref{inv1}). 

Equations (\ref{inv3}, \ref{inv5}) show that the boundary conditions 
in the vertex of the graph are determined (up to a $C$-factor) by the scattering data or, 
more precisely, by the value $S_0$ of the scattering matrix at fixed $k_0$. 
This description \cite{H2} of the boundary conditions has the advantage of being 
non-degenerate\footnote{We thank the referee for this observation.}. 
Kirchhoff's rule and scale invariance are imposed in these coordinates 
by requiring 
\be 
S_0 {\bf v} = {\bf v}\, , \qquad 
S_0^* = S_0 \, . 
\label{inv7}
\ee 
The classification of the critical points is therefore equivalent to the classification 
of the $S_0$-matrices satisfying (\ref{inv2}) and (\ref{inv7}).

\sect{Vertex operators}

We turn now to the construction of vertex operators on the graph $\Gamma$. 
Although such operators exist \cite{Mintchev:2005rz} for the general 
boundary condition (\ref{bc}), at a scale invariant point they have 
simpler and more remarkable structure. For this reason we focus here 
on the {\it scale invariant} case, introducing first the right and left chiral fields
\be 
\ph_{i, R} (t-x)=\ph(t,x,i)+\phd(t,x,i)\, , \qquad 
\ph_{i,L}(t+x)=\ph(t,x,i)-\phd(t,x,i)\, . 
\label{rlbasis}
\ee 
Inserting (\ref{sol1},\ref{sol2}) in (\ref{rlbasis}) one gets 
\be
\ph_{i,R}(\xi ) = \int_{0}^{\infty} \frac{\rd k}{\pi \sqrt
{2k}}
\left[a^{\ast i}(k) \e^{\ri k\xi} +
a_i (k) \e^{-\ri k\xi}\right ] \,  , 
\label{pir}
\ee
\be
\ph_{i,L} (\xi) = \int_{0}^{\infty} 
\frac{\rd k}{\pi \sqrt {2k}} 
\left[a^{\ast i}(-k) \e^{\ri k\xi} +
a_i (-k) \e^{-\ri k\xi}\right ] \, . 
\label{pil}
\ee 
The algebraic features of the chiral fields (\ref{pir},\ref{pil}) can be easily deduced from 
(\ref{ccr1}-\ref{ccr3}). One finds 
\be
[\ph_{i_1,R}(\xi_1)\, ,\, \ph_{i_2,R}(\xi_2)] =  
-i\varepsilon (\xi_{12})\, \delta_{i_1}^{i_2} \, , 
\label{comm1}
\ee
\be 
[\ph_{i_1,L}(\xi_1)\, ,\, \ph_{i_2,L}(\xi_2)] 
=  -i\varepsilon (\xi_{12})\, \delta_{i_1}^{i_2} \, ,
\label{comm2}
\ee
\be
[\ph_{i_1,R}(\xi_1)\, ,\, \ph_{i_2,L}(\xi_2)] =   
-i\varepsilon (\xi_{12})\, S_{i_1}^{i_2} \, , 
\label{comm3}
\ee 
where $\xi_{12}\equiv \xi_1-\xi_2$ and we have used that $S$ is $k$-independent at a scale invariant point. 

The chiral charges associated with (\ref{pir},\ref{pil}) are 
\be 
Q_{i,Z} = \frac{1}{4} \int_{-\infty}^{\infty} \rd \xi \, \prt_\xi \ph_{i,Z} (\xi)\, , 
\qquad Z=R,\, L\, , 
\label{lrcharges}
\ee 
and satisfy the commutation relations 
\be 
[Q_{i_1,R}\, ,\, \ph_{i_2,R}(\xi)] = 
[Q_{i_1,L}\, ,\, \ph_{i_2,L}(\xi)] = -\frac{\ri}{2} \delta_{i_1}^{i_2}\, , 
\label{qcomm1}
\ee 
\be 
[Q_{i_1,R}\, ,\, \ph_{i_2,L}(\xi)] = 
[Q_{i_1,L}\, ,\, \ph_{i_2,R}(\xi)] = -\frac{\ri}{2} S_{i_1}^{i_2} \, , 
\label{qcomm2}
\ee 
\be 
[Q_{i_1,Z_1}\, ,\, Q_{i_2,Z_2}]=0\, . 
\label{qcomm3}
\ee  

At this point we are ready to introduce a family of vertex operators 
parametrized by $\zeta = (\sigma , \tau) \in \RR^2$ and defined by 
\be  
v(t,x,i;\zeta) = z_i\, q(i;\zeta) 
:\exp\left \{\ri \sqrt{\pi}\left [\sigma \ph_{i,R}(t-x) + 
\tau \ph_{i,L}(t+x)\right]\right \}: \, , 
\label{vertex1}
\ee 
where the value of the normalization constant $z_i\in \RR$ will be fixed later on and 
\be 
q(i;\zeta)= \exp\left [\ri \sqrt{\pi}\left (\sigma Q_{i,R} -\tau Q_{i,L}\right )\right ]\, .  
\label{qfact}
\ee 
The exchange properties of $v(t,x,i;\zeta )$ 
determine their statistics. A standard calculation shows that 
\be 
v(t_1,x_1,i_1;\zeta_1 ) v(t_2,x_2,i_2;\zeta_2 ) = 
\R (t_{12},x_1,i_1,x_2,i_2; \zeta_1,\zeta_2 )\, 
v(t_2,x_2,i_2;\zeta_2 ) v(t_1,x_1,i_1;\zeta_1 ) \, , 
\label{exch}
\ee
the exchange factor $\R$ being a c-number. 
The statistics of $v(t,x,i;\zeta )$ is determined by the 
value of $\R$ at space-like separation $t_{12}^2-x_{12}^2 <0$. 
By means of (\ref{comm1}-\ref{comm3}) and (\ref{qcomm1}-\ref{qcomm3}) one finds  
\be 
\R(t_{12}, x_1,i_1, x_2,i_2;\zeta_1,\zeta_2 ){\Big \vert_{t_{12}^2-x_{12}^2 <0}} = 
\e^{-\ri \pi (\sigma_1 \sigma_2 -\tau_1 \tau_2) \varepsilon (x_{12}) \delta_{i_1}^{i_2}}
\, . 
\label{exchf1}
\ee 
Therefore $v(t,x,i;\zeta )$ obey anyon (abelian braid) statistics with parameter 
\be 
\vartheta (\zeta_1,\zeta_2)=  \sigma_1 \sigma_2 -\tau_1 \tau_2 \, ,
\label{statpar}
\ee 
when localized at the same wedge $E_i$. Otherwise, $v(t,x,i;\zeta )$ commute. 

In the next section we shall bosonize two-component Dirac fermions on the 
graph $\Gamma$. For this purpose we take any $\zeta = (\sigma, \tau)$ with $\sigma \not= \pm \tau$ and set 
\be 
\zeta^\prime = (\tau, \sigma)\, . 
\label{zprime}
\ee 
Then we define 
\be 
V(t,x,i;\zeta ) = \eta_i\, v(t,x,i;\zeta )\, , 
\qquad V(t,x,i;\zeta^\prime ) = \eta_i^\prime\, v(t,x,i;\zeta^\prime )\, ,
\label{vertex2}
\ee 
where $\{\eta_i, \eta_i^\prime\}$ are the so called Klein factors, which generate an associative algebra $\cal K$ 
with identity $\bf 1$ and satisfy the anticommutation relation 
\be 
\eta_{i_1} \eta_{i_2} + \eta_{i_2} \eta_{i_1} = 2 \delta_{i_1i_2} {\bf 1}\ , \quad  
\eta_{i_1}^\prime \eta_{i_2}^\prime + \eta_{i_2}^\prime \eta_{i_1}^\prime = 2 \delta_{i_1i_2} {\bf 1}\, ,  \quad 
\eta_{i_1} \eta_{i_2}^\prime + \eta_{i_2}^\prime \eta_{i_1} = 0 \, . 
\label{klein}
\ee 
{}From (\ref{exchf1}, \ref{klein}) one deduces that (\ref{vertex2}) and their Hermitian conjugates 
obey Fermi statistics provided that 
\be 
 \sigma^2 -\tau^2 = 2k+1\, , \qquad   k\in \ZZ \, . 
 \label{fermi}
 \ee 
 In what follows we denote by ${\cal V}_\zeta$ the vertex algebra generated by 
 \be 
 \{V(t,x,i;\zeta )\, , V(t,x,i;\zeta^\prime )\, ,  V^*(t,x,i;\zeta )\, , V^*(t,x,i;\zeta^\prime )\} 
 \label{valg}
 \ee 
 {}for fixed $\zeta \in \RR^2$. Without loss of generality one can take $\sigma >0$, 
 which is assumed throughout the paper. The algebra ${\cal V}_\zeta$ is the basic tool for bosonization.

\bigskip
\sect{Bosonization on star graphs} 

It is worth stressing that all the results of sections 2-4 
hold on a general algebraic level and do not refer to a specific representation of the 
algebras $\A$ and $\cal K$. In this sense they are universal. For the physical applications 
we have in mind, we fix below the Fock representation of $\A$ and define a simple 
representation of the algebra of Klein factors $\cal K$.

\subsection{Correlation functions} 

The basic correlators are 
\be 
\langle  \ph_{i_1,Z_1} (\xi_1)  \ph_{i_2,Z_2} (\xi_2)\rangle = 
(\ph_{i_1,Z_1} (\xi_1) \Omega\, ,\,  \ph_{i_2,Z_2} (\xi_2) \Omega ) \, , 
\label{cf1}
\ee 
where $\Omega$ and $(\cdot\, ,\, \cdot )$ are the vacuum state and the scalar product in the 
Fock representation \cite{Liguori:1996xr} of $\A$. 
Using the exchange relations (\ref{ccr1}-\ref{ccr3}) and the fact that 
$a_i(k)$ annihilate $\Omega$ one easily derives 
\be 
\langle \ph_{i_1,R}(\xi_1) \ph_{i_2,R}(\xi_2)\rangle = 
\langle \ph_{i_1,L}(\xi_1) \ph_{i_2,L}(\xi_2)\rangle =
\delta_{i_1}^{i_2}\, u(\mu \xi_{12}) \, , 
\label{cf2}
\ee 
where 
\be 
u(\mu \xi )= \int_{0}^{\infty} \frac{\rd k}{\pi }(k^{-1})_{\mu}  \e^{-\ri k\xi}  
\label{distr1}
\ee 
with $(k^{-1})_{\mu}$ defined by \cite{Liguori:1997vd} 
\be 
(k^{-1})_{\mu} = \frac{\rd}{\rd k} \ln\frac{k\, \e^{\gamma_E}}{\mu} \, .  
\label{distr2}
\ee 
The derivative here is understood in the sense of distributions, $\gamma_E$ is Euler's constant and 
$\mu>0$ is a free parameter with dimension of mass having a well-known 
infrared origin. The integral (\ref{distr1}), computed by means of the representation (\ref{distr2}), gives 
\be 
u(\mu \xi)=-\frac{1}{\pi} \ln (\mu |\xi|) -\frac{i}{2}\varepsilon (\xi) = 
-\frac{1}{\pi} \ln (i\mu \xi + \epsilon )\, , \qquad \epsilon > 0\, . 
\label{log}
\ee 

Analogously, for the mixed $L-R$ correlators one finds 
\be 
\langle \ph_{i_1,R}(\xi_1) \ph_{i_2,L}(\xi_2)\rangle = 
\int_{0}^{\infty} \frac{\rd k}{\pi }(k^{-1})_{\mu}  \e^{-\ri k\xi_{12}} S_{i_1}^{i_2}(k)\, ,  
\label{cf3}
\ee
\be 
\langle \ph_{i_1,L}(\xi_1) \ph_{i_2,R}(\xi_2)\rangle = 
\int_{0}^{\infty} \frac{\rd k}{\pi }(k^{-1})_{\mu}  \e^{-\ri k\xi_{12}} S_{i_1}^{i_2}(-k)\, .   
\label{cf4}
\ee
As expected, (\ref{cf3},\ref{cf4}) keep track of the interaction at the junction and  
are quite complicated for the general $S$-matrix (\ref{S1}). 
There is however a remarkable simplification at any scale invariant point, because $S$ is constant. 
One has in fact  
\be 
\langle \ph_{i_1,R}(\xi_1) \ph_{i_2,L}(\xi_2)\rangle = 
\langle \ph_{i_1,L}(\xi_1) \ph_{i_2,R}(\xi_2)\rangle = S_{i_1}^{i_2}\, u(\mu \xi_{12}) \, .  
\label{cf6}
\ee 

In order to compute the correlation functions of the vertex operators (\ref{vertex2}), 
we need also a representation of the algebra $\cal K$ of Klein factors. We adopt the one defined by 
the two-point correlators 
\be 
\langle \eta_{i_1} \eta_{i_2} \rangle = 
\langle \eta_{i_1}^\prime \eta_{i_2}^\prime \rangle = \kappa_{i_1i_2}=
\left\{\begin{array}{cc}
\; \; 1 \, ,
& \quad \mbox{$i_1\leq i_2$}\, ,\\[1ex]
-1\, ,
& \quad \mbox{$i_1>i_2$}\, , \\[1ex]
\end{array} \right.
\label{klein2}
\ee 
\be 
\langle \eta_{i_1} \eta_{i_2}^\prime \rangle = 
-\langle \eta_{i_2}^\prime \eta_{i_1} \rangle = \kappa_{i_1i_2} \, .
\label{klein3}
\ee 
Denoting by $\eta_i^\natural$ any of the factors $\eta_i$ and $\eta_i^\prime$, 
the $n$-point functions are given by 
\be 
\langle \eta_{i_1}^\natural \cdots \eta_{i_n}^\natural \rangle =
\left\{\begin{array}{cc}
\; \; 0 \, ,
& \quad \mbox{$n=2k+1$}\, ,\\[1ex]
\sum_{p \in {\cal P}_{2k}} \E_p 
\langle \eta_{p_1}^\natural \eta_{p_2}^\natural \rangle  \cdots \langle \eta_{p_{2k-1}}^\natural \eta_{p_{2k}}^\natural \rangle\, ,
& \quad \mbox{$n=2k$}\, , \\[1ex]
\end{array} \right.
\label{klein4}
\ee 
where the sum runs over all permutations ${\cal P}_{2k}$ of the numbers $1,2,...,2k$ 
and $\E_p$ is the parity\footnote{In other words $\E_p=1$ and $\E_p=-1$ for 
even and odd permutations respectively.} of the permutation $p$. 

We turn now to the vertex correlation functions, defining first the concept of {\it physical} 
vertex operator $V_{\rm ph}(t,x,i;\zeta)$. It is defined by selecting among all vertex 
correlation functions those which are invariant under the shift transformations (\ref{shiftsymm}), or equivalently, by the selection rules  
\be 
\sum_{j=1}^n \sigma_j= \sum_{j=1}^n \tau_j =0\, ,
\label{selection}
\ee
which are the counterpart of the ``neutrality" condition in the Coulomb gas approach 
to conformal field theory in 1+1 dimensions. 
Now $V_{\rm ph}(t,x,i;\zeta)$ are defined via their vacuum expectation values given by  
\bea 
\langle V_{\rm ph}(t_1,x_1,i_1;\zeta_1)\cdots V_{\rm ph} (t_n,x_n,i_n;\zeta_n) \rangle 
\qquad \qquad \qquad \nonumber \\
=\left\{\begin{array}{cc}
\langle V(t_1,x_1,i_1;\zeta_1)\cdots  V(t_n,x_n,i_n;\zeta_n) \rangle\, ,
& \quad \mbox{(\ref{selection}) holds}\, , \\[1ex]
\; \; 0 \, ,
& \quad \mbox{(\ref{selection}) is violated}\, .\\[1ex]
\end{array} \right.
\label{physcorr}
\eea  

Let us concentrate now on the vertex algebra ${\cal V}_\zeta$. A standard computation 
shows that the non-trivial two-point vertex functions are   
\bea  
\langle V_{\rm ph}(t_1,x_1,i_1;\zeta) V^*_{\rm ph}(t_2,x_2,i_2;\zeta) \rangle = 
\qquad \qquad \qquad \qquad \nonumber \\
z_{i_1}z_{i_2}
 \mu^{-[(\sigma^2+\tau^2)\delta_{i_1}^{i_2} +2\sigma \tau S_{i_1}^{i_2}]} 
 \kappa_{i_1i_2}
\left [\frac{1}{\ri (t_{12}-x_{12})+ \epsilon}\right ]^{\sigma^2 \delta_{i_1}^{i_2}} 
\left [\frac{1}{\ri (t_{12}+x_{12})+ \epsilon}\right ]^{\tau^2 \delta_{i_1}^{i_2}}
\quad \nonumber \\
\left [\frac{1}{\ri (t_{12}-{\widetilde x}_{12})+ \epsilon}\right ]^{\sigma \tau S_{i_1}^{i_2}}
\left [\frac{1}{\ri (t_{12}+{\widetilde x}_{12})+ \epsilon}\right ]^{\sigma \tau S_{i_1}^{i_2}} ,\, \, 
\label{cf7}
\eea
\bea  
\langle V_{\rm ph}(t_1,x_1,i_1;\zeta^\prime ) V^*_{\rm ph}(t_2,x_2,i_2;\zeta^\prime ) \rangle = 
\qquad \qquad \qquad \qquad \nonumber \\
z_{i_1}z_{i_2}
 \mu^{-[(\sigma^2+\tau^2)\delta_{i_1}^{i_2} +2\sigma \tau S_{i_1}^{i_2}]} 
 \kappa_{i_1i_2}
\left [\frac{1}{\ri (t_{12}-x_{12})+ \epsilon}\right ]^{\tau^2 \delta_{i_1}^{i_2}} 
\left [\frac{1}{\ri (t_{12}+x_{12})+ \epsilon}\right ]^{\sigma^2 \delta_{i_1}^{i_2}}
\quad \nonumber \\
\left [\frac{1}{\ri (t_{12}-{\widetilde x}_{12})+ \epsilon}\right ]^{\sigma \tau S_{i_1}^{i_2}}
\left [\frac{1}{\ri (t_{12}+{\widetilde x}_{12})+ \epsilon}\right ]^{\sigma \tau S_{i_1}^{i_2}} ,\, \, 
\label{cf8}
\eea
with ${\widetilde x}_{12}=x_1+x_2$. These equations suggest to take the normalization factor 
\be 
z_i = \mu^{\frac{1}{2}(\sigma^2+\tau^2 + 2\sigma \tau S_i^i)} \, . 
\label{z}
\ee 
In this way the vertex correlators (\ref{cf7}, \ref{cf8}) are $\mu$-independent, when 
localized on the same edge. 

Performing in (\ref{cf7}) the scaling transformation $t_i \mapsto \varrho t_i$ 
and $x_i\mapsto \varrho x_i$ one obtains 
\be 
\langle V_{\rm ph}(\varrho t_1,\varrho x_1,i_1;\zeta) V_{\rm ph}^*(\varrho t_2,\varrho x_2,i_2;\zeta) \rangle = 
\varrho^{-D_{i_1}^{i_2}}\, \langle V_{\rm ph}(t_1,x_1,i_1;\zeta) V_{\rm ph}^*(t_2,x_2,i_2;\zeta) \rangle \, , 
\label{sctransf}
\ee
where 
\be 
D = (\sigma^2 +\tau^2)\II_n + 2\sigma \tau S\, . 
\label{msc}
\ee 
The scaling dimensions $d_i$ are determined by the eigenvalues of the matrix $D$. 
Diagonalizing (\ref{msc}) one finds 
\be 
d_i = \frac{1}{2}(\sigma^2 + \tau^2) + \sigma \tau s_i \, , \qquad i=1,...,n \, ,  
\label{dimensions0}
\ee 
where $s_i$ are the eigenvalues of $S$. From (\ref{unit2}) one infers that in the scale 
invariant case $s_i=\pm 1$, which implies 
\be 
d_i = \frac{1}{2}(\sigma + s_i\tau)^2 \geq 0 \, .   
\label{dimensions}
\ee 
Recalling that the same vertex operator on the line $\RR$ has dimension 
\be 
d_{\rm line} =  \frac{1}{2}(\sigma^2 + \tau^2) \, ,   
\label{dimline}
\ee 
we see that the interaction at the junction affects the scaling dimensions.  
More precisely, the deviation of $d_i$ from $d_{\rm line}$ reflects the interaction 
between the left and right chiral fields at the junction.

\subsection{Bosonization} 

The massless Dirac equation on the star graph $\Gamma$ is 
\be 
(\gamma_t \prt_t - \gamma_x \prt_x)\psi (t,x,i) = 0 \, , 
\label{deq} 
\ee 
where
\be 
\psi (t,x,i)=\pmatrix{ \psi_1(t,x,i) \cr \psi_2(t,x,i) \cr}\, ,  
\qquad 
\gamma_t=\gamma^t=\pmatrix{ 0 & 1 \cr 1 & 0 \cr}\, , \qquad
\gamma_x=-\gamma^x=\pmatrix{ 0 & 1 \cr -1 & 0 \cr} \, .
\label{gamma}
\ee 
The standard vector and axial currents are 
\be 
j_\nu (t,x,i) = \overline \psi (t,x,i) \gamma_\nu \psi (t,x,i) \, , \quad  
\qquad j_\nu^5 (t,x,i) = \overline \psi (t,x,i) \gamma_\nu \gamma^5 \psi (t,x,i) \,  , 
\label{currents} 
\ee
with $\overline \psi \equiv \psi^\ast \gamma_t $ and 
$\gamma^5 \equiv -\gamma_t\gamma_x $. From eq.(\ref{deq}) it follows 
that both $j_\nu $ and $j_\nu^5$ are conserved.  
Moreover, the $\gamma^5$-identities $\gamma_t\gamma^5 = -
\gamma_x $ and 
$\gamma_x \gamma^5 = - \gamma_t $ imply the relations 
\be
j_t^5 = - j_x \, ,\quad  \qquad j_x^5 = -j_t \, . 
\label{currel}
\ee

Our goal now is to quantize (\ref{deq}) using the vertex algebra ${\cal V}_\zeta$ and to express 
the currents in terms of $\{\ph,\, \phd \}$. For this purpose we set $\zeta =(\sigma >0,0)$ 
and define 
\be
\psi_1 (t,x,i) = \frac{1}{\sqrt {2\pi}}V_{\rm ph}(t,x,i;\zeta )\, ,  
\qquad 
\psi_2 (t,x,i) =\frac{1}{\sqrt {2\pi}} V_{\rm ph}(t,x,i;\zeta^\prime ) \, . 
\label{psi} 
\ee 
One easily verifies that (\ref{psi}) satisfy the Dirac equation (\ref{deq}) 
and obey Fermi statistics if  
\be 
\sigma^2 = 2k+1\, , \qquad k\in \NN\, ,   
\label{fcond}
\ee 
{}For the equal-time anticommutators of $\psi_a$ 
and $\psi^{*a}$ ($a = 1,2$) one finds 
\be 
\{\psi_{a_1} (0,x_1,i_1)\, ,\, \psi^{* a_2} (0,x_2,i_2)\} = 
\frac{(-1)^k}{(2k)!} \delta^{(2k)}(x_{12})\delta_{i_2}^{i_1} 
\delta_{a_1}^{a_2} \, , 
\label{kcomm}
\ee
showing that the conventional canonical fermions are obtained for $k=0$. 
We analyze below the general case $k\in \NN$. 

The next step is to construct the quantum currents (\ref{currents}). We 
adopt the point-splitting procedure, considering the limit  
\be
j_\nu (t,x,i) = {1\over 2} \lim_{\epsilon \to +0} Z(\epsilon) 
\left [\, \overline \psi (t,x,i)\gamma_\nu \psi (t, x+\epsilon,i ) + 
\overline \psi (t, x+\epsilon,i) \gamma_\nu \psi (t,x,i)\, \right ]\, ,
\label{pointsplit1}  
\ee
where $Z(\epsilon)$ implements the renormalization. 
The basic general formula for 
evaluating (\ref{pointsplit1}) is obtained by normal ordering the product 
$V^\ast_{\rm ph} (t,x+\epsilon,i;\zeta ) V_{\rm ph}(t,x,i;\zeta )$. After some algebra 
\cite{Mintchev:2005rz}, one derives the renormalization constant     
\be 
Z(\epsilon ) = 
{-\pi \epsilon^{\sigma^2 -1}\over \sigma \sin \left ({\pi\over 2}\sigma^2\right ) }\, , 
\label{rc1} 
\ee 
which leads, performing the limit in (\ref{pointsplit1}), 
to the conserved current 
\be 
j_\nu (t,x,i) = {\sqrt \pi} \prt_\nu \ph (t,x,i) = {\sqrt \pi}\, k_\nu (t,x,i)  \, .  
\label{jph} 
\ee 
Thus one recovers on the star graph $\Gamma$ the same type of relation as in conventional 
bosonization \cite{bos}. 

In analogy with (\ref{pointsplit1}) we introduce the axial current by  
\be 
j_\nu^5 (t,x,i) = 
{1\over 2} \lim_{\epsilon \to +0} Z(\epsilon) 
\left [\, \overline \psi (t,x,i)\gamma_\nu \gamma^5 \psi (t, x+\epsilon,i ) + 
\overline \psi (t, x+\epsilon,i ) \gamma_\nu \gamma^5 \psi (t,x,i)\, \right ]\, , 
\label{pointsplit2}  
\ee 
The vector current result and the $\gamma^5$-identities directly imply that 
the limit in the right hand side of (\ref{pointsplit2}) exists and 
\be 
j^5_\nu (t,x,i) = {\sqrt \pi}\, \prt_\nu \phd (t,x,i) = {\sqrt \pi}\, \kt_\nu (t,x.i) \, . 
\label{jphd} 
\ee 
Eq. (\ref{eqm2}) shows that the relations (\ref{currel}) are respected after bosonization as well. 

Another essential aspect is the Kirchhoff rule. 
The results of section 2.1 imply that the currents $j_\nu$ and $j^5_\nu$ cannot satisfy simultaneously 
Kirchhoff's rule. In what follows we assume (\ref{kir2}), which guaranties the conservation 
of the vector charge. The case when $j_\nu^5$ satisfies Kirchhoff's rule can be analyzed 
along the same lines. 

Eqs.(\ref{jph},\ref{jphd}) imply that the boundary conditions on 
$\psi $ at the vertex of the star graph are most conveniently formulated in terms of the 
currents, which are the simplest observables of the fermion field. 
Combining  (\ref{bc}) with (\ref{jph},\ref{jphd}) one obtains 
\be
\sum_{j=1}^n A_i^j  \int_{+0}^{\infty} \rd x \, j_x(t,x,j) = \sum_{j=1}^n B_i^j\,  j_x(t,0,j) \, .   
\label{fbc}
\ee 
where we have used that $\lim_{x\to \infty} \ph (t,x,i) = 0$ holds on the subspace 
generated by the currents $j_\nu$ and $j_\nu^5$.

\subsection{Conductance} 

In order to explore the conductance properties 
of a quantum wire with the form of a star graph, 
we study here the linear response of the current $j_\nu$ to a classical external potential 
$A_\nu$ minimally coupled to $\psi$, namely 
\be 
\gamma^\nu [\prt_\nu + \ri A_\nu(t,x,i)] \psi (t,x,i) = 0 \, .
\label{deqa} 
\ee 
The main step is to extend the bosonization procedure of the previous section, 
deriving an action in terms of $\ph$ and $A_\nu$ which implements the 
dynamics defined by (\ref{deqa}) and preserves its invariance under the local 
gauge transformations 
\bea 
\psi(t,x,i)&\mapsto & \e^{\ri \Lambda(t,x,i)}\psi(t,x,i)\, ,
\label{gauge1}\\
A_\nu(t,x,i)&\mapsto & A_\nu(t,x,i)-\partial_\nu\Lambda(t,x,i)\, . 
\label{gauge2}
\eea 
For this purpose we first observe that according to (\ref{psi}) the transformation (\ref{gauge1}) 
is implemented by the shift 
\be
\ph(t,x,i)\mapsto \ph(t,x,i)+
\frac{1}{\sigma \sqrt{\pi}}\Lambda(t,x,i)\, ,
\qquad \phd(t,x,i)\mapsto \phd(t,x,i)\, ,
\ee 
where $\sigma $ satisfies (\ref{fcond}).   
Therefore, switching on $A_\nu$ the bosonized current becomes by gauge invariance  
\be
j_\nu(t,x,i)={\sqrt \pi}\partial_\nu\ph(t,x,i) +\frac{1}{\sigma}A_\nu (t,x,i) \, , 
\label{curra}
\ee 
The action, which ensures the conservation of (\ref{curra}), is 
\be 
I[\ph, A_\nu]=\frac{1}{2}\sum_{i=1}^n \int_{-\infty}^\infty \rd t \int_0^\infty \rd x 
\left [ \prt^\nu \ph \prt_\nu \ph + \frac{2}{\sigma {\sqrt \pi}} \prt^\nu \ph A_\nu + 
\frac{1}{\sigma^2\pi}A^\nu A_\nu \right ](t,x,i) \, . 
\label{act}
\ee
In fact, varying (\ref{act}) with respect to $\ph$, one obtains 
\be 
\prt^\nu \prt_\nu \ph (t,x,i) + 
\frac{1}{\sigma {\sqrt \pi} } \prt^\nu A_\nu (t,x,i) = 0\, , 
\label{curracons}
\ee 
which is precisely the conservation of (\ref{curra}). The interaction Hamiltonian 
associated to (\ref{act}) is 
\be
H_{\rm int}(t)= \sum_{i=1}^{n} 
\int_0^{\infty}\rd x \left [\frac{1}{\sigma \sqrt{\pi}}  \partial_x\ph A_x  
- \frac{1}{2\sigma^2 \pi}  A^\nu A_\nu \right ](t,x,i) \, .
\label{h}
\ee 

At this stage we are ready to derive the expectation value $\langle j_x(t,x,i)\rangle_{A_\nu}$ in 
the external field $A_\nu$. Keeping in mind that $A_\nu$ is classical, 
the linear response theory \cite{FW} gives 
\bea
\langle j_x(t,x,i)\rangle_{A_\nu} =  
\langle j_x(t,x,i) \rangle +\ri\int_{-\infty}^{t} \rd\tau
\langle [H_{\rm int}(\tau)\, ,\, j_x(t,x,i)]\rangle =\nonumber \\
\frac{1}{\sigma} A_x(t,x,i)+
\frac{\ri }{\sigma} \sum_{j=1}^{n}\int_{-\infty}^{t} \rd\tau 
\int_0^\infty \rd y A_y(\tau,y,j)
\langle [\partial_y\ph(\tau,y,j)\, ,\, \partial_x\ph(t,x,i)]\rangle \, .
\label{lrt1}
\eea 
Let us consider now a uniform electric field  
$E(t,i)=\partial_t A_x(t,i)$ in the Weyl gauge $A_t=0$. 
Using (\ref{comm1}-\ref{comm3}) one derives from (\ref{lrt1}) 
\be 
\langle j_x(t,0,i)\rangle_{A_x}=
\frac{1}{2\sigma }\sum_{j=1}^n \left ( \delta_i^j-S_i^j \right ) A_x(t,j) \, , 
\label{lrt2}
\ee
which leads to the conductance tensor 
\be
G_i^j =\frac{1}{2\sigma }\left (\delta_i^j-S_i^j \right ) \, .
\label{conduc1}
\ee 
Because of (\ref{kir2}) $G_i^j$ satisfies Kirchhoff's rule 
\be
\sum_{j=1}^n G_i^j =0 \, , \qquad i=1,...,n\, , 
\label{kircond}
\ee
representing an useful check of the whole derivation. 

By means of the same technique and conventions one obtains for a single infinite wire 
(without any junction) 
\be 
G_{\rm line} = \frac{1}{2\sigma }\, , 
\label{line}
\ee 
which allows one to rewrite (\ref{conduc1}) in the form 
\be
G_i^j =G_{\rm line} \left (\delta_i^j-S_i^j \right ) \, . 
\label{cond2}
\ee 
The unitarity of the $S$-matrix implies $|S_i^i | \leq 1$, leading to the simple bound 
\be 
0\leq G_i^i \leq 2 G_{{\rm line}} \, ,  
\label{bound}
\ee 
where we have used that $\sigma$ and therefore $G_{\rm line}$ are positive. 
Another constraint on the diagonal elements of $G$ is obtained from 
(\ref{tr1}), which gives the sum rule 
\be 
{\rm Tr}\, G = 2p\, G_{{\rm line}}\, , 
\label{tr2}
\ee
where $p$ is the rank of the matrix $A$ entering the boundary condition (\ref{bc}). 

It is instructive at this point to consider some examples. For the critical points (\ref{sc1}) 
one has for instance (no summation over $i$) 
\be
G_i^i = 2\left (1-\frac{1}{n}\right ) G_{\rm line}\mathop{\longrightarrow}_{n\to\infty} 2G_{\rm line}\, , 
\label{cond4}
\ee
showing an enhancement with respect to $G_{\rm line}$ for $n\geq 3$. This remarkable phenomenon 
has been discovered in \cite{NFLL}, where it has been interpreted as a result of the so called 
Andreev reflection \cite{A}. Notice also that the enhancement is growing with the number of wires $n$. 

A closer examination of the case $n=3$ reveals another interesting feature. At the critical point 
(\ref{sc2}) all three edges have the same enhanced conductance 
\be 
G_1^1 = G_2^2 = G_3^3 = \frac{4}{3} G_{\rm line} \, . 
\label{ccond4}
\ee
The situation is quite different however for the family of critical points (\ref{sc4}). 
In that case the sum rule (\ref{tr2}) gives 
\be 
G_1^1(\alpha ) + G_2^2(\alpha ) +  G_3^3 (\alpha )= 2 G_{\rm line} \, , 
\qquad \alpha \in \RR\, .  
\label{cond5}
\ee
We have plotted the conductance of all three edges in Fig. 2. 
\begin{figure}[h]
\begin{center}
\includegraphics[scale=0.5]{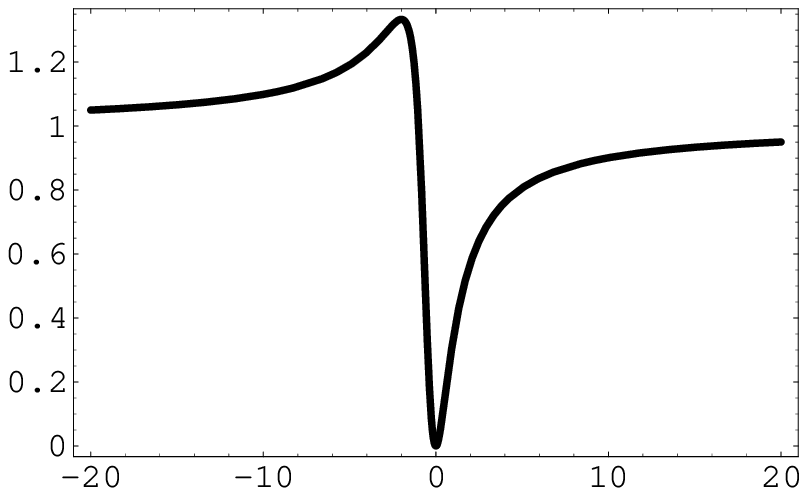}
\hskip 0.7 truecm 
\includegraphics[scale=0.5]{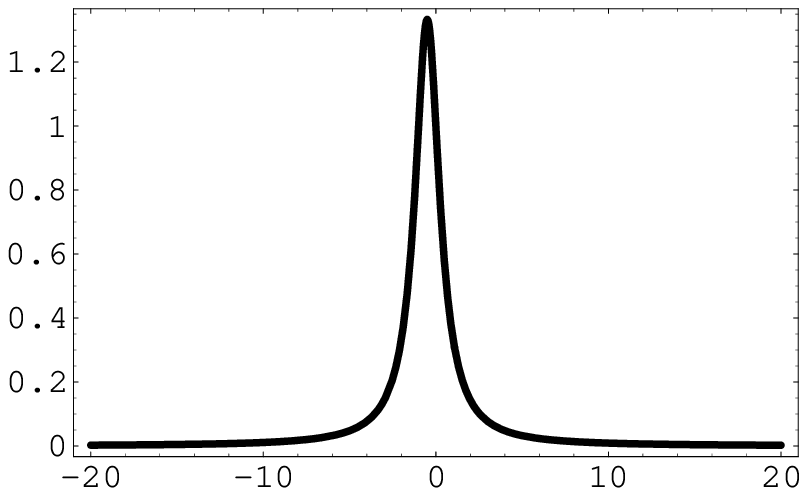}
\hskip 0.7 truecm 
\includegraphics[scale=0.5]{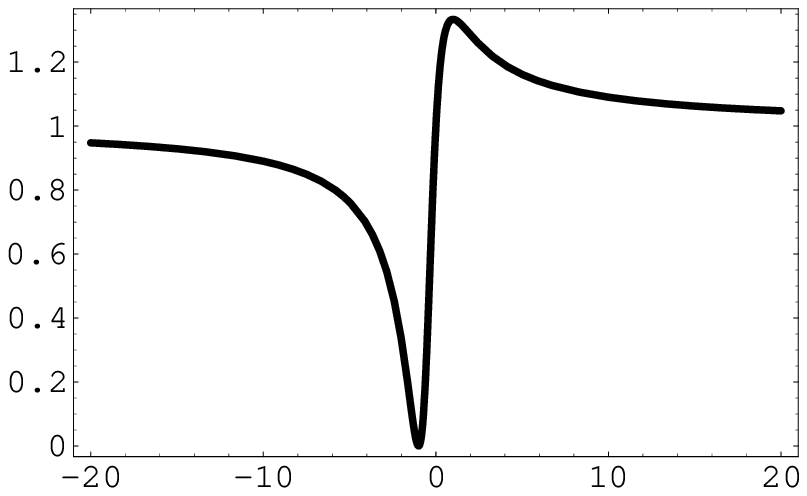}
\end{center}
\centerline{Figure 2: Plots of $G_1^1 (\alpha)$ (left), $G_2^2 (\alpha)$ (center) and 
$G_3^3 (\alpha)$ (right) for $G_{{\rm line}}=1$.} 
\end{figure} 
Besides of the domains of enhancement 
(the maxima of the curves are exactly at $4G_{\rm line}/3$), 
there are domains where the conductance is depressed 
and points (the minima), where it actually vanishes. This characteristic property of the 
new critical points, discovered in the present paper, may have interesting 
applications. Let us observe finally that for $n=3$ one has by inspection 
\be 
0\leq G_i^i \big \vert_{n=3} \leq \frac{4}{3} G_{{\rm line}} \, , 
\label{ccond5}
\ee 
which is sharper than the unitarity bound (\ref{bound}). 

It is easy to see that the conductance for the axial current $j_\nu^5$ 
is obtained from (\ref{cond2}) by the substitution $S_i^j \longmapsto -S_i^j$. 

Eq. (\ref{cond2}) represents a nice and universal formula which holds at any critical point 
and determines the conductance in terms of the interaction at the junction. 
An analogous expression has been derived in \cite{Bellazzini:2006jb} for 
another physical observable - the Casimir 
energy density $\E_{\rm C}(x,i)$. One has 
\be 
\E_{\rm C}(x,i)=-\frac{1}{8\pi x^2}\, S_i^i \, , 
\label{casimir}
\ee 
establishing a direct relation between conductance and Casimir effect on quantum wires. 
Eq. (\ref{casimir}) implies that in the case of enhanced conductance ($S_i^i<0$), 
the energy density $\E_{\rm C}(x,i)$ gives rise to a repulsive 
Casimir force.\footnote{There is recently a growing interest 
\cite{Barton, Fulling:2006ts} in this phenomenon.} 
If instead the conductance is depressed, the Casimir force is attractive. 

We would like to mention in conclusion 
that the above technique applies to the derivation of the conductance 
away from the critical points as well. Adopting the general correlators (\ref{cf2}, \ref{cf3}, \ref{cf4}) 
one obtains 
\be
G_i^j(\omega) = G_{{\rm line}} \left [\delta_i^j-S_i^j(\omega)\right ] \, ,
\label{cond6}
\ee 
where $\omega$ is the frequency of the Fourier transform ${\widehat A}_x(\omega,i)$ of the 
external field $A_x(t,i)$ applied to the system. 

In the next section we shall introduce non-trivial scale invariant bulk interactions 
and investigate their impact on the conductance. 

\bigskip
\sect{The massless Thirring model} 

The classical dynamics of the massless Thirring model \cite{Thirr} is governed 
by the equation of motion 
\be 
i(\gamma_t \prt_t - \gamma_x \prt_x)\Psi (t,x,i) = 
\lambda\left [\gamma_t J_t(t,x,i) - 
\gamma_x J_x (t,x,i) \right ] \Psi (t,x,i) 
\, , \quad x\not=0\, , 
\label{theqm}
\ee 
where $\lambda>0 $ is the coupling constant and $J_\nu$ is 
the conserved current 
\be 
J_\nu (t,x,i) = \overline \Psi (t,x,i) \gamma_\nu \Psi (t,x,i) \, .  
\label{thcurr}
\ee 
The system is scale invariant and can be quantized by means of the vertex algebra ${\cal V}_\zeta$ 
with $\zeta =(\sigma>0,\tau )$. We set 
\be
\Psi_1 (t,x,i) = \frac{1}{\sqrt {2\pi}}V_{\rm ph}(t,x,i;\zeta )\, , \qquad 
\Psi_2 (t,x,i) = \frac{1}{\sqrt {2\pi}}V_{\rm ph}(t,x;\zeta^\prime ) \, ,
\label{psit}
\ee 
where $\sigma$ and $\tau$ satisfy (\ref{fermi}) in order to have Fermi statistics. 

The quantum current $J_\nu $ is constructed 
in analogy with (\ref{pointsplit1}), setting  
\be
J_\nu (t,x,i) = {1\over 2} \lim_{\epsilon \to +0} Z (x,i;\epsilon) 
\left [\, \overline \Psi (t,x,i)\gamma_\nu \Psi (t, x+\epsilon, i) + 
\overline \Psi (t, x+\epsilon, i) \gamma_\nu \Psi (t, x, i)\, \right ]\, . 
\label{pointsplit3}  
\ee 
Using the short distance expansion of the product $V_{\rm ph}^*(t,x+\epsilon,i;\zeta )V_{\rm ph}(t,x,i;\zeta )$ 
one finds
\be 
Z(x,i;\epsilon) = \frac{-\pi \epsilon^{\sigma^2 + \tau^2 -1}(2x)^{\sigma \tau S_i^i}}
{(\sigma-\tau)\sin\left[\frac{\pi}{2}(\sigma^2-\tau^2)\right]}\, , 
\label{ZZ}
\ee
and 
\be 
J_\nu (t,x) = {\sqrt \pi }\, \prt_\nu \ph (t,x) \, .  
\label{thvcurrent}
\ee
The $x$ and $i$-dependence of $Z$ are not surprising because translations along the edges and 
their permutations are not symmetries in general. Because of (\ref{thvcurrent}) the quantum 
equation of motion takes the form 
\be
i(\gamma_t \prt_t - \gamma_x \prt_x)\Psi (t,x,i) = 
\lambda {\sqrt \pi } : \left (\gamma_t \prt_t \ph - 
\gamma_x \prt_x \ph \right ) \Psi : (t,x,i)  \, .  
\label{qeqm}
\ee
Now, using the explicit form (\ref{psit}) of $\Psi$, one easily 
verifies that (\ref{qeqm}) is satisfied provided that 
\be 
\tau = -\frac{1}{2}\lambda \, .
\label{condeqm}
\ee 
Combining eq. (\ref{fermi}) and eq. (\ref{condeqm}), 
we obtain for $\sigma$  
\be 
\sigma = \sqrt {{\lambda^2\over 4} + (2k+1) } \, , 
\label{sol} 
\ee
where 
\be 
k \in \ZZ\, , \qquad k \geq -\frac{\lambda^2+4}{8} \, ,
\label{condsol} 
\ee
ensuring that $\sigma \in \RR$. The freedom associated with 
$k$ is present also in the Thirring model on the line. 
Since Lorentz invariance is preserved there, it is natural to 
require that the Lorentz spin of $\Psi $ takes 
the canonical value $1/2$, which fixes $k=0$. 

{}For deriving the conductance of the Thirring model one can apply the result of section 5.3. 
One introduces an external field $A_\nu$ minimally coupled to $\Psi$ and observes 
that the local gauge transformations of $\Psi$ are now implemented by the shift 
\be
\ph(t,x,i)\mapsto \ph(t,x,i)+
\frac{1}{(\sigma + \tau) \sqrt{\pi}}\Lambda(t,x,i)\, ,
\qquad \phd(t,x,i)\mapsto \phd(t,x,i)\, .
\ee
Performing the same steps as in section 5.3, one gets 
\be 
G_i^j = \frac{1}{2(\sigma +\tau)} \left (\delta_i^j - S_i^j \right ) \, ,  
\label{thcond1}
\ee 
where the bulk interaction is captured by the factor 
\be 
\frac{1}{2(\sigma +\tau)}= \frac{1}{\sqrt{\lambda^2+4(2k+1)} - \lambda}\, . 
\label{thcond2}
\ee 
As expected, eqs. (\ref{thcond1}, \ref{thcond2}) reproduce (\ref{conduc1}) for $\lambda=0$. 
We see also that the effect of the current-current bulk interaction on the conductance is 
an overall $\lambda$-dependent renormalization of (\ref{conduc1}). Since 
the bulk interaction preserves scale invariance, all critical points of the 
Thirring model on a star graph are those described in section 3. From (\ref{dimensions}) the 
scaling dimensions at a given critical point are 
\be 
d_i=\frac{1}{4}\left [\sqrt{\lambda^2+4(2k+1)}-s_i \lambda \right ]^2 \, , 
\label{Thdim}
\ee 
where $s_i$ are the eigenvalues of the related $S$-matrix.

\bigskip
\sect{Remarks and conclusions} 

The interest in various aspects of quantum field 
theory and critical phenomena on star graphs is not new and is growing recently. Among others, 
we already mentioned the very inspiring papers \cite{KF}-\cite{Oshikawa:2005fh} 
on this subject. The interaction at the junction of the wires is implemented 
there by means of a Lagrangian ${\cal L}_{\rm int}(t,0,i)$, 
localized at the vertex $x=0$ of the star graph. After bosonization, ${\cal L}_{\rm int}(t,0,i)$ involves typically 
exponential interactions of the bulk scalar fields $\ph$ and $\phd$. 
The models arising this way are investigated by various methods including conventional 
perturbation theory, perturbative conformal field theory, 
instanton gas expansion, functional renormalization group and so on. In this context 
some critical points and the interpolating renormalization group flows have been determined. 

In the present work we propose an alternative strategy, based on the point-like character 
of the vertex interactions, the powerful theory of self-adjoint extension of Hermitian operators 
on a graph \cite{Kostrykin:1998gz}-\cite{KSch} and the algebraic technique 
\cite{Liguori:1996xr}-\cite{Mintchev:2005rz} for 
dealing with defects. Within the new approach we establish the
complete  classification of the critical points of the massless 
scalar field $\ph$ on a star graph with any number of edges. 
The main idea is to study the bulk dynamics of $\ph$, isolating first the vertex 
of the graph. The vertex and the associated interaction are recovered afterwards 
constructing all possible unitarity preserving extensions of the bulk theory to the whole graph. 
We derived (sect. 3) in this way new families of critical points with peculiar 
conductance properties (sect. 5). Besides domains of enhancement, we discover also 
critical points in which the conductance is depressed. It turns out that the Casimir force 
has a different sign in these two regimes: enhancement of the conductance corresponds 
to a repulsion and depression to attraction. It is worth stressing that 
our framework applies also away from criticality. Indeed, 
we were able to construct explicitly the renormalization 
group flows for a broad class of boundary conditions and associate 
with them a simple action $I[\ph]$ which, instead of being exponential, is quadratic in $\ph$. 
Non-trivial bulk interactions have been treated by this technique as well. 
We focused (sect. 6) on the Thirring model, which is a ``relativistic" 
generalization of the Tomonaga-Luttinger model. 

The relation of our framework to boundary conformal field theory 
(BCFT) \cite{Cardy:1984bb}-\cite{Cardy:2004hm} is another interesting issue. The key point for 
understanding the interplay between BCFT and scale invariance on quantum graphs 
is the Kirchhoff rule (\ref{kirteta}) for the energy-momentum tensor $\theta_{tx}$. For $n=1$ this 
rule implies the absence of momentum flow across the boundary and is precisely 
the starting point of conventional BCFT. In fact, for $n=1$ our approach reproduces the results of BCFT 
for the scalar field $\ph$. The relative scale-invariant boundary conditions 
are the Dirichlet and Newmann conditions, which correspond to {\it complete reflection} from the boundary. 
A new phenomenon takes place for $n\geq2$. In fact, besides reflection one can have in this case 
also {\it non-trivial transmission} between the different edges. In other words, $\theta_{tx}(t,0,i)$ need 
not to vanish separately for any $i$, still respecting (\ref{kirteta}). The spectrum of scale-invariant 
boundary conditions for $n\geq2$ is richer (sect 3), the scale dimensions are affected by the
interaction  at the junction (sect. 5) and one is naturally led to a generalization of the $n=1$
BCFT. Besides the isolated  critical points, this generalization involves multi-parameter
families of such points,  which shed new light on the critical properties of quantum wires. For
this reason we believe that conformal field theory, defined on star graphs by Kirchhoff's rule
(\ref{kirteta}) independently of a particular set  of fundamental fields (Lagrangians), needs
further attention. 

It will be interesting to explore also the concept of integrability on quantum graphs. 
A first step in this direction is the analysis \cite{Bellazzini:2006jb} of the nonlinear
Schr\"odinger equation on a star graph. The extension of the above framework to  
finite temperature and/or generic quantum graphs represents also a challenging open problem, 
whose solution will surely help for better understanding the physics of quantum wires. 

\bigskip 
\bigskip 
\bigskip

\noindent{\bf Acknowledgments} 
\bigskip 

\noindent 
Work supported in part by the TMR Network EUCLID: "Integrable models and applications: from strings to condensed matter", contract number HPRN-CT-2002-00325.
\bigskip
\bigskip

\noindent {\Large\bf Appendix} 
\bigskip

As explained in section 3.4, there exist two families of $n=4$ critical points, 
each one depending on two parameters $\alpha_1, \alpha_2 \in \RR$. For $p=1$ the $S$-matrix is defined by: 
\bea 
S_1^1 &=& \frac{1}{\Delta_1}(\alpha_1 +\alpha_1 ^2+\alpha_2 +\alpha_1  \alpha_2 +\alpha_2 ^2 )\, ,\nonumber \\
S_2^2 &=& \frac{1}{\Delta_1}(1+\alpha_1 +\alpha_1 ^2+\alpha_2 +\alpha_1  \alpha_2 )\, ,\nonumber \\
S_3^3 &=& \frac{1}{\Delta_1}(1+\alpha_1 +\alpha_2 +\alpha_1  \alpha_2 +\alpha_2^2 )\, ,\nonumber \\
S_4^4 &=& -\frac{1}{\Delta_1}(\alpha_1+\alpha_2+\alpha_1 \alpha_2)\, ,\nonumber \\
\nonumber
\eea 
\bea
S_1^2&=&-\frac{1}{\Delta_1}\alpha_2\, ,\quad \quad \; \; 
S_1^3=-\frac{1}{\Delta_1}\alpha_1\, ,\qquad \qquad \qquad  
S_1^4=\frac{1}{\Delta_1}(1+\alpha_1+\alpha_2)\, ,
\nonumber \\
S_2^3&=&-\frac{1}{\Delta_1}\alpha_1 \alpha_2\, ,\quad \; \, 
S_2^4=\frac{1}{\Delta_1}\alpha_2(1+\alpha_1+\alpha_2)\, ,\quad 
S_3^4=\frac{1}{\Delta_1}\alpha_1(1+\alpha_1+\alpha_2)\, , 
\nonumber 
\eea
\medskip 
with $\Delta_1 ={1+\alpha_1 +\alpha_1 ^2+\alpha_2 +\alpha_1  \alpha_2 +\alpha_2 ^2}$. The remaining entries 
are recovered by symmetry. 

Analogously, the $S$-matrix of the family corresponding to $p=2$ is given by: 
\bea 
S_1^1 &=& \frac{1}{\Delta_2}[3 \alpha_1 ^2+2 \alpha_1  (1-\alpha_2 )-(1+\alpha_2 )^2 ]\, ,\nonumber \\
S_2^2 &=& \frac{1}{\Delta_2}[-1-\alpha_1 ^2+2 \alpha_2 +3 \alpha_2 ^2-2 \alpha_1 (1+\alpha_2 )]\, ,\nonumber \\
S_3^3 &=& \frac{1}{\Delta_2}[3-\alpha_1^2+2\alpha_2-\alpha_2^2+2\alpha_1(1+\alpha_2)]\, ,\nonumber \\
S_4^4 &=& -\frac{1}{\Delta_2}[\alpha_1 ^2+2 \alpha_1  (1-\alpha_2 )+(1+\alpha_2 )^2]\, ,\nonumber \\
\nonumber
\eea 
\bea
S_1^2&=&\frac{2}{\Delta_2}(1+\alpha_1 +\alpha_2 +2 \alpha_1 \alpha_2 )\, ,\qquad 
S_1^3=\frac{2}{\Delta_2}[\alpha_2  (1+\alpha_2 )-\alpha_1  (2+\alpha_2 )]\, ,\qquad 
\nonumber \\
S_1^4&=&\frac{2}{\Delta_2}(1+\alpha_1 -\alpha_1  \alpha_2 +\alpha_2 ^2)\, ,\qquad \, \; 
S_2^3=\frac{2}{\Delta_2}(\alpha_1 +\alpha_1 ^2-2 \alpha_2 -\alpha_1  \alpha_2 )\, ,
\nonumber \\
S_2^4&=&\frac{2}{\Delta_2}(1+\alpha_1 ^2+\alpha_2 -\alpha_1  \alpha_2 )\, ,\qquad \, \; 
S_3^4=\frac{2}{\Delta_2}(\alpha_1 +\alpha_1 ^2+\alpha_2 +\alpha_2 ^2)\, , 
\nonumber 
\eea
\medskip 
where $\Delta_2 ={3+3 \alpha_1 ^2+2 \alpha_1  (1-\alpha_2)+2 \alpha_2 +3 \alpha_2 ^2}$. 

As a check on the above $S$-matrices one can verify the validity of (\ref{kir2}) and (\ref{tr1}).

\end{document}